\documentclass[fleqn,usenatbib]{mnras}

\usepackage{newtxtext,newtxmath}

\usepackage[T1]{fontenc}
\usepackage{ae,aecompl}

\usepackage{graphicx}	
\usepackage{amsmath}	
\usepackage{amssymb}	

\title[High-Resolution RT Modelling of M33]{High-Resolution Radiative Transfer Modelling of M33}

\author[T. G. Williams et al.]
{Thomas G. Williams,$^{1}$\thanks{Email: thomas.williams@astro.cf.ac.uk}
Maarten Baes,$^{2}$
Ilse De Looze,$^{2,3}$
Monica Rela\~{n}o,$^{4,5}$\newauthor
Matthew W. L. Smith,$^{1}$
Sam Verstocken,$^{2}$ and
S\'ebastien Viaene$^{2,6}$
\\
${1}$ School of Physics \& Astronomy, Cardiff University, Queens Buildings, The Parade, Cardiff, CF24 3AA, UK\\
${2}$ Sterrenkundig Observatorium, Universiteit Gent, Krijgslaan 281, B-9000 Gent, Belgium\\
${3}$ Department of Physics and Astronomy, University College London, Gower Street, London WC1E 6BT, UK\\
${4}$ Departmento de F\'isica Te\'orica y del Cosmos, Facultad de Ciencias, Universidad de Granada, E-18071 Granada, Spain\\
${5}$ Instituto Universitario Carlos I de F\'isica Te\'orica y Computacional, Universidad de Granada, E-18071 Granada, Spain\\
${6}$ Centre for Astrophysics Research, University of Hertfordshire, College Lane, Hatfield, AL10 9AB, UK
}

\date{Accepted XXX. Received YYY; in original form ZZZ}

\pubyear{2019}

\begin{document}
\label{firstpage}
\pagerange{\pageref{firstpage}--\pageref{lastpage}}
\maketitle

\begin{abstract} 
In this work, we characterise the contributions from both ongoing star formation and the ambient radiation field in Local Group galaxy M33, as well as estimate the scale of the local dust-energy balance (i.e. the scale at which the dust is re-emitting starlight generated in that same region) in this galaxy through high-resolution radiative transfer (RT) modelling, with defined stellar and dust geometries. We have characterised the spectral energy distribution (SED) of M33 from UV to sub-mm wavelengths, at a spatial scale of 100\,pc. We constructed input maps of the various stellar and dust geometries for use in the RT modelling. By modifying our dust mix (fewer very small carbon grains and a lower silicate-to-carbon ratio as compared to the Milky Way), we can much better fit the sub-mm dust continuum. Using this new dust composition, we find that we are able to well reproduce the observed SED of M33 using our adopted model. In terms of stellar attenuation by dust, we find a reasonably strong, broad UV bump, as well as significant systematic differences in the amount of dust attenuation when compared to standard SED modelling. We also find discrepancies in the residuals of the spiral arms versus the diffuse interstellar medium (ISM), indicating a difference in properties between these two regimes. The dust emission is dominated by heating due to the young stellar populations at all wavelengths ($\sim$80\% at 10\,\micron\, to $\sim$50\% at 1\,mm). We find that the local dust-energy balance is restored at spatial scales greater than around 1.5\,kpc.
\end{abstract}




\section{Introduction}\label{sec:introduction}

Despite only contributing around 1\% of the mass of the interstellar medium (ISM) of a galaxy, dust absorbs, scatters, and reprocesses around 30\% of the starlight in star-forming galaxies \citep[e.g.][]{2002Popescu,2016Viaene}. An understanding of the processes governing the interactions of stars and dust is, therefore, essential to understanding how galaxies evolve, their dust properties, and extracting important intrinsic parameters such as the star formation rate (SFR) and initial mass function (IMF). The starlight absorbed in UV and optical is re-emitted by the dust at far-infrared (FIR) and sub-mm wavelengths. Assuming only absorption of light from younger stars, the total infrared (TIR) luminosity can therefore be used as a proxy for star-formation \citep[see, e.g.][]{2011Murphy}. Alternatively, by understanding the (wavelength-dependent) amount of dust attenuation, wavebands that suffer from attenuation can be corrected using some combination of dust measurements \citep[e.g.][]{2008Leroy,2011Hao}, or by assuming some dust model \citep[e.g.][]{2000CharlotFall}. 

One method of modelling the light from a galaxy is by fitting an SED across these wavelengths, often using a large library of models, and several tools are available for this purpose \citep[e.g.][]{2008DaCunha,2009Noll,2019Boquien,2016Chevallard}. However, these tools assume a local dust-energy balance (i.e. that the dust emission per unit area comes from light originating from stars in that same area), which may be unsuitable for modelling sub-kpc regions \citep{2015Boquien,2018Smith}. These tools also neglect the 3D geometry of a galaxy, and do not consider the propagation vectors of photons through this medium. For a complete study of the interactions of the dust and stellar components of a galaxy, 3D radiative transfer (RT) models are required, which take into account this 3D geometry and are not beholden to a per-pixel local dust-energy balance. There are a number of codes available for this purpose (see \citealt{2013Steinacker} for a review of these, as well as an overview of the RT mathematics). Due to the complexity of the RT calculations, and the fact that these calculations are both non-linear and non-local, most of these codes make use of Monte Carlo (MC) or ray-tracing techniques. Unlike traditional SED fitting, RT is computationally very expensive, and thus faces its own series of challenges, such as loss of information due to projection effects, and the difficulty of applying traditional solution algorithms to these problems.

Previous work in this area has tended to focus on ``simpler'', better-behaved galaxies such as edge-on (or nearly edge-on) spirals \citep[e.g.][]{2001Misiriotis,2008Bianchi,2010Baes,2012aDeLooze,2012bDeLooze,2014DeGeyter,2015DeGeyter,2016Mosenkov,2018Mosenkov}. Galaxies at lower inclinations have also been modelled, including the spiral galaxy M51 \citep{2014DeLooze} and very nearby galaxy M31 \citep{2017aViaene}, finding significant variations in dust heating by old and young stellar populations; these works also find that the relative contributions to dust heating are both wavelength- and position-dependent. A large step in increasing the complexity of these simulations was employed by \cite{2014DeLooze}, using observed images to describe the distribution of stars and dust. A framework for modelling face-on galaxies is currently in development by Verstocken et al. (in prep.), which will be applied to a number of the DustPedia \citep{2017Davies} galaxies. Different approaches to RT modelling of galaxies have also been employed -- in particular, taking an axisymmetric approach, \cite{2017Popescu} have produced an RT model for the Milky Way (MW).

In this work, we perform a high-resolution RT simulation of the third massive spiral galaxy in our Local Group, M33 (the Triangulum Galaxy). Being the third largest spiral on the sky, smaller only than our own MW and M31, and with a close proximity of 840kpc \citep{1991MadoreFreedman}, it is an excellent target of choice for high-resolution observations. M33 has been mapped across many wavelengths with a variety of observatories. Due to the wealth of high-resolution data, this galaxy is therefore naturally suited for detailed RT simulations. M33 has a roughly half-solar metallicity ($12+\log(\text{O/H})=8.36\pm0.04$, \citealt{2008RosolowskySimon}), and a shallow metallicity gradient. This lower metallicity makes M33 a very different environment to M31 and the MW, more analogous to younger, higher redshift galaxies. As the RT model is 3D, the data is necessarily deprojected, and a third dimension modelled, but with a moderate inclination of 56\degr \citep{1994ReganVogel}, the deprojection degeneracies are not as pronounced as in M31. M33 has significant star-formation across its disc \citep{2004Heyer}, with SFRs between 0.2\,$\mathrm{M}_\odot$ yr$^{-1}$ and 0.45\,$\mathrm{M}_\odot$ yr$^{-1}$, depending on the SFR tracer used \citep{2007Verley,2018Williams}. Given its relatively small size ($R_{25}$ = 30.8\arcmin, $\sim$7.4\,kpc, \citealt{2003Paturel}), this means that M33 has a much higher star formation efficiency than other Local Group galaxies (with a gas depletion timescale of 1.6-3.2$\times10^8$\,yr; \citealt{2007Gardan}). Because of this active star formation, we may expect a higher contribution to the overall dust heating by younger stellar populations, but it is important not to neglect the effect of dust heating by older stellar populations.

Earlier RT studies of M33 have focussed on the nucleus \citep{1999Gordon}, and in modelling the global SED \citep{2016Hermelo}. \cite{1999Gordon} modelled only the ultraviolet to near-infrared (UV-NIR) SED of this nucleus, finding evidence of strong dust attenuation. \cite{2016Hermelo} applied the RT model of \cite{2011Popescu} which uses a series of axisymmetric models to describe the various geometries of the galaxy, and produced a global SED from UV-sub-mm wavelengths. The goal of this study was to investigate the ``sub-mm excess'', which appears to be present in many low-metallicity environments \citep[e.g.][]{2010Bot,2011Galametz,2013Kirkpatrick,2013RemyRuyer}. The main conclusion of this work was that likely, the sub-mm excess could be accounted for by modifying the dust grain composition. Our study seeks to build on these previous works, studying the attenuation of M33 on a global level with a richer data-set than \cite{1999Gordon}, as well as to modify the dust grain properties in our input model to better fit the data, and to use input geometries based on observables to produce a resolved study of many of these properties.

The layout of this paper is as follows: we present an overview of the dataset we use in this work (Sect. \ref{sec:data}), before an overview of the setup of our RT model (Sect. \ref{sec:model}). We then fit this model to the observed SED of M33 (Sect. \ref{sec:model_fit}), before investigating some of the global and resolved properties of M33 (Sect. \ref{sec:results}). Finally, we summarise our main conclusions in Sect. \ref{sec:conclusions}.

\section{Data}\label{sec:data}

The data used in this work is largely the same as in \cite{2018Williams}, and we refer the reader to that work for a more detailed description. A brief description is given here. Both FUV and NUV data was obtained \citep{2005Thilker} by the Galaxy Evolution Explorer (GALEX; \citealt{2005Martin}). In the optical, Sloan Digital Sky Survey (SDSS; \citealt{2000York}) data was mosaicked together using primary frames from the SDSS DR13 \citep{2015Alam}, using {\sc montage}\footnote{\url{http://montage.ipac.caltech.edu}}. A 3 square degree mosaic was created for all of the u, g, r, i, and z bands, to allow us to accurately model background variations. We also make use of H$\alpha$ data \citep{2000HoopesWalterbos}, which was not included in the previous work of \cite{2018Williams}. This map has a pixel size of around 2\,arcsec, and covers a total field-of-view of 1.75\,deg$^2$. This map has also been continuum-subtracted. Corrections for contamination from [N {\sc ii}] emission have not been carried out, although it is estimated that a maximum of 5\% of the flux could result from [N {\sc ii}] emission in any region of the galaxy \citep{2000HoopesWalterbos}.

For near- and mid-infrared, we make use of \textit{Spitzer} and Wide-field Infrared Survey Explorer (WISE; \citealt{2010Wright}) data. The former of these was taken as part of the Local Volume Legacy (LVL, \citealt{2009Dale}) survey, with Infrared Array Camera \citep[IRAC,][]{2004Fazio} data at 3.6, 4.5, 5.8 and 8\,\micron \,and Multiband Imaging Photometer \citep[MIPS,][]{2004Rieke} data at 24 and 70\,\micron. The latter covers a similar wavelength range to the former, with data at 3.4, 4.6, 12, and 22\,\micron, and has been mosaicked together from the ALLWISE data release, which includes both the WISE cryogenic and NEOWISE \citep{2011Mainzer} post-cryogenic phase.

Far-infrared and sub-mm data was obtained from the \textit{Herschel} space observatory \citep{2010Pilbratt} and the Submillimetre Common-User Bolometer Array 2 \citep[SCUBA-2,][]{2013Holland} on the James Clerk Maxwell Telescope (JCMT). As part of the HerM33es \citep{2010Kramer} open time key project, M33 was mapped by both the Photodetector Array Camera and Spectrometer \citep[PACS,][]{2010Poglitsch} at 100 and 160\,\micron, and the Spectral and Photometric Imaging REceiver \citep[SPIRE,][]{2010Griffin} at 250, 350, and 500\,\micron. The SCUBA-2 data is at 450 and 850\,\micron, and we use the technique presented in Smith et al. (in prep.), to maintain the high resolution offered by SCUBA-2, but add back in the large-scale structure that is lost in the data reduction process. Details of this SCUBA-2 data reduction are given in \cite{2019Williams}. We note that this SCUBA-2 data does not cover the entirety of M33, so for global flux values we use the SPIRE 500\,\micron\ and Planck 850\,\micron\ fluxes. In total, our dataset covers almost 4 orders of magnitude in wavelength, from 1516\,\AA\ to 850\,\micron.

\begin{table*}
 \caption{Overview of parameters in the model. Although most parameters are fixed, for free parameters we indicate the parameter search range and the wavelength these luminosities are normalised at. All luminosities are given in $L_\odot$ at the normalisation wavelength. The dust mass is given in M$_\odot$. The error in each parameter is calculated by sampling the likelihood distributions, and is quoted as half the bin width if the sampled error is smaller than a single bin.}
 \label{tab:skirt_parameters}
 \centering
 \begin{tabular}{cccc}
  \hline
  \hline
  Component & Parameter & Value & Best Fit Luminosity/Mass ($L_\odot$/M$_\odot$)\\
  \hline				
   & 2D Geometry & IRAC 3.6\,\micron & \\
  Old Stellar Disc & Total Luminosity & $0.4 - 4 \times 10^8$ (3.6\,\micron) & $(2.8^{+1}_{-0.5}) \times 10^8$ \\
   & Vertical Scale Height & 200\,pc & \\
  \hline
   & 2D Geometry & GALEX FUV$^1$ & \\
  Non-Ionizing Stellar Disc & Total Luminosity & $0.8 - 5 \times 10^9$ (0.15\,\micron) & $(1.7 \pm 0.5) \times 10^9$\\
   & Vertical Scale Height & 100pc & \\
  \hline
   & 2D Geometry & H$\alpha$ + 24\,\micron$^2$ & \\
  Ionizing Stars & Total Luminosity & $0.3 - 3.3\times 10^7$ (0.66\,\micron) & $(3.3 \pm 1.5) \times 10^7$ \\
   & Vertical Scale Height & 50\,pc & \\
  \hline
   & 2D Geometry & {\sc magphys} Dust Mass Map$^3$ & \\
  Dust & Total Dust Mass & $2.5-7 \times 10^6 \,\textrm{M}_\odot$ & $(3.6 \pm 0.6) \times 10^6$ \\
   & Vertical Scale Height & 100\,pc & \\
  \hline
  \multicolumn{4}{p{2\columnwidth}}{$^1$Corrected for attenuation and diffuse emission. $^2$Corrected for diffuse stellar emission. $^3$Obtained from pixel-by-pixel {\sc magphys} fitting.}
 \end{tabular}
\end{table*}

For each of these images, we have performed a number of steps to make this diverse dataset homogeneous. For frames in which foreground star emission is present, we masked this using UV colours \citep{2008Leroy}. We performed a Galactic extinction correction using the prescription of \cite{2011SchlaflyFinkbeiner}, using extinction values calculated for the central position of M33, although we note that due to M33's large angular extent, this correction varies across the face of the disc, which is taken into account in our treatment of the uncertainties. However, for the GALEX bands, which are most affected by this variation in extinction correction, the difference in flux is maximally $\sim$3\%, which is negligible when combined with the other errors considered \citep[see][for more details on this error analysis]{2018Williams}. We then convolved all of the data to our worst working resolution, the SPIRE 350\,\micron\ beam (which has a FWHM of 25\,arcsec, corresponding to 100\,pc at the distance of M33). This data is then regridded to pixels of 25\,arcsec, so that they can be considered statistically independent.

With this dataset homogenized, we performed pixel-by-pixel SED fitting for the $\sim$19000 pixels within a radius of 60\,arcmin $\times$ 70\,arcmin, using the SED fitting tool {\sc magphys} \citep{2008DaCunha}, and we refer readers to this work for details on the {\sc magphys} model details. This allows us to calculate a number of intrinsic quantities of the galaxy, and provides both an attenuated and unattenuated SED for each pixel, with the attenuation following the model of \cite{2000CharlotFall}. This modelling technique has previously been employed by \cite{2016Viaene} for their modelling of M31, and means that we can make immediate comparison with this earlier work. We also note that \cite{2014Viaene} and \cite{2018Williams} find that {\sc magphys} produces similar results to more conventional measures of, e.g., dust mass and SFR with observational data at resolutions of 130 and 100\,pc, respectively. \cite{2018Smith} find statistically acceptable fits to many key galaxy properties, when compared to simulated data at resolutions of 200\,pc to 25\,kpc.

\section{The 3D Model}\label{sec:model}

For the radiative transfer simulations, we make use of {\sc skirt}\footnote{\url{skirt.ugent.be}} \citep{2003Baes,2015CampsBaes}, a publicly available, Monte Carlo RT code. This code was originally developed to investigate the effects of dust extinction on the photometry and kinematics of galaxies, but has developed to accurately model the absorption, scattering and emission of starlight by dust. It has also been tested against the major benchmarks published that the code is applicable to \citep[e.g][]{2015Camps}. {\sc skirt} can accept an arbitrary number of components to model, where each of these components are defined by a 3D geometry, an intrinsic spectrum, and a normalisation of this spectrum (either at a given wavelength, or a bolometric luminosity. This code allows for panchromatic RT simulations, using a wide variety of geometry models and optional modifiers for these geometries \citep{2015BaesCamps}. It also provides a number of options for efficient dust grids \citep{2014Saftly}, for which we use a binary tree adaptive grid method. This means that we can effectively increase the resolution in dense regions (such as spiral arms), while minimising the computational cost of this increased resolution. The code can also model stochastically heated dust grains \citep{2015Camps}. It is also provided with parallelisation, to allow these computationally expensive simulations to run efficiently \citep{2017Verstocken}. Finally, it allows for the input of a 2D FITS image as a geometry, which was first employed by \cite{2014DeLooze} in the grand-design spiral galaxy M51, and which we use to define our various geometries in this work. {\sc skirt} deprojects and derotates this image given an inclination and position angle, and assumes that the distribution of pixel values in this input image corresponds to the density in a linear way. It then scales this map to a total density provided when setting up the geometry, and conserves total flux during deprojection. This 2D model is then given extra dimensionality by assuming an exponential profile with a provided vertical scale height (which will vary for each input geometry). 

To make our notation consistent throughout this work, but comparable to earlier studies, we refer to flux densities using the symbol $S$, and luminosities as $L$. Fractions of these quantities will be referred to with the symbol $\mathcal{F}$.

\subsection{Model Components}\label{sec:model_components}

A typical galaxy model setup for RT simulations composes of a bulge and thick disc containing old stars, with a thin star-forming disc containing dust and young stars \citep[e.g.][]{1999Xilouris,2000Popescu}. We use this model with one alteration -- M33 does not appear to have a bulge, at least in the traditional sense \citep{1992Bothun}. This claim is somewhat controversial, but for the purposes of our work we treat M33 as bulge-less. This means that we assume all of the old stars reside within the same exponential disc, rather than a population at the centre extending much further above the plane of the galaxy. We use three stellar components in our model: the first represents the old stellar populations (stars of ages around $\sim$8\,Gyr; Sect. \ref{sec:old_stellar_disc}). The second stellar component consists of the young stars that are UV bright but dissociated from their birth clouds, and have ages around 100\,Myr (Sect. \ref{sec:young_stellar_disc}). Our final stellar component are the young stars still present in their birth clouds, and producing hard, ionizing radiation (Sect. \ref{sec:ionizing_stars}). We refer to the combination of these young non-ionizing and ionizing stellar populations as ``young'' throughout this work. We also provide a map of the dust mass surface density, which traces the dust distribution within the galaxy (Sect. \ref{sec:dust_model}). The details of this modelling approach are based on Verstocken et al. (in prep.).

For each component, we specify an input geometry, a particular SED type and a luminosity normalisation. Along with this, we provide an input FITS image, where we have truncated the disc to 1.2\,$R_{25}$ and set to 0 any pixels that correspond to those that have signal-to-noise (S/N) < 5 in the SPIRE 350\,\micron\ map (the map that defines our working resolution). For each stellar component, we specify a metallicity of 12 + log(O/H) = 8.36, corresponding to the central metallicity of M33 \citep{2008Rosolowsky}. Whether M33 has a radial gradient in its metallicity is a topic of contention. Whilst \cite{2008Rosolowsky} find a slight radial gradient, \cite{2011Bresolin} find no such significant gradient. In either case, the practical effect this would have on the form of the SED is minor. Finally, in all cases for the geometries we assume a position angle of 22.5\degr \citep{1959deVaucouleurs} and an inclination of 56\degr \citep{1994ReganVogel}. A summary of the major parameters of our model are given in Table \ref{tab:skirt_parameters}.

\subsubsection{Old Stellar Disc}\label{sec:old_stellar_disc}

The geometry of our old stellar component is set by the IRAC 3.6\,\micron\, image, which is generally considered to be a pure tracer of stellar mass \citep[e.g.][]{2010Zhu}. In our initial testing, we found significant contribution from the young stellar populations at this wavelength, and so using the {\sc magphys} star formation history (SFH) we make a first-order correction to separate out the contribution of these younger populations from the total luminosity. We note that as we leave the luminosity of each stellar component as a free parameter, this is only needed for a first guess. There may also be a contribution at this wavelength from hot dust heated by the young stars, but leaving the stellar luminosities as free parameters in our fitting will effectively account for this. Also, in practice the contribution from young populations is likely position-dependent, but given the coarse nature of the {\sc magphys} SFH, performing robust corrections of this nature is beyond the scope of this work. We normalise the luminosity of the old stellar disc at 3.6\,\micron.

For a panchromatic simulation, we require an emitted luminosity at each wavelength for each component. We do this by taking a template SED, and matching the observed emission to this. In the case of this old stellar population, we make use of the \cite{2003BruzualCharlot} simple stellar populations (SSPs), at an age of 8\,Gyr, which we assume is the average age of these older stars. Finally, to make this geometry 3D, we assume an exponential profile for the disc, characterised by a vertical scale height. Generally, the scale height of the old stellar populations is taken to be 1/8.86 the scale length \citep{2013DeGeyter}. In M33 this scale length is $1.82\pm0.02$\,kpc \citep{2015Kam}, giving us a scale height of $\sim$200\,pc.

\subsubsection{Non-Ionizing Stellar Disc}\label{sec:young_stellar_disc}

The first of our young stellar populations are the stars of age $\sim$100\,Myr, which are UV bright but unable to ionize hydrogen. These stars are only attenuated by dust in the diffuse ISM, and so suffer much less from dust attenuation than those stars in the birth clouds. We used as our initial input geometry the GALEX FUV image, which traces unobscured star formation over the last 10-100\,Myr \citep{1999Meurer}. We calculated an unattenuated flux for each pixel by convolving the unattenuated {\sc magphys} SED with the GALEX FUV filter response, which effectively corrects for the effects of dust attenuation.

Although the FUV is dominated by these young stars, there can also be a significant amount of UV flux from more diffuse, older, stellar populations, which we correct for using the prescription of \cite{2008Leroy}:
\begin{equation}
S_\text{FUV, young} = S_\text{FUV, unatten} - \alpha_\text{FUV} S_{3.6},
\end{equation}
where $\alpha_\text{FUV} = 3\times10^{-3}$, and $S_x$ is in Jy. Given that \cite{2008Leroy} do not correct the 3.6\,\micron\, flux for young stars when calculating this factor, we use the uncorrected 3.6\,\micron \,flux. As these young stars are expected to reside within a thinner disc than the old stars, we adopt a scale height of 100\,pc, half that of the old stellar component, and we normalise the extinction-corrected luminosity at the FUV wavelength.

\subsubsection{Ionizing Stars}\label{sec:ionizing_stars}

Our final stellar component consists of very young (<10\,Myr) stars that are still embedded in their birth clouds and produce hard, ionizing radiation. This radiation is difficult to trace directly, but can be inferred from H$\alpha$ emission, and dust grains heated to high temperatures. To create a map of the ionizing radiation, we used a continuum-subtracted H$\alpha$ map, and combined this with a map of the hot dust. The hot dust is traced via the 24\micron\ emission, which, much like the FUV map, we corrected for a diffuse stellar component:
\begin{equation}
S_\text{24, ion} = S_\text{24} - \alpha_\text{24} S_\text{3.6}.
\end{equation}
These fluxes are again in Jy, and $\alpha_\text{24}$ was determined by \cite{2008Leroy} to be 0.1. This factor was calculated from a sample of nearby galaxies, but appears to be robust throughout the sample (see the discussion in their appendix D.2.4), and thus should be applicable to M33. The input geometry for the ionizing map is then 
\begin{equation}
S_{{\rm H}\alpha{\rm , ion}} = S_{{\rm H}\alpha} + 0.031 S_\text{24, ion},
\end{equation}
where fluxes are in ergs/s \citep{2007Calzetti}.

The SED we used for this input geometry was the MAPPINGS III \citep{2008Groves} nebular modelling code, and we refer readers to that work for definitions of the various parameters of the model. Generally, we adopt the same parameters for this SED as \cite{2014DeLooze}, with a compactness of $\log C=5.5$, and a surrounding ISM pressure of $1\times10^{12}\,\text{K m}^{-3}$. However, we choose a slightly lower cloud covering factor of 0.1, half that of \cite{2014DeLooze}, which we found in initial testing to be a slightly better fit to the data. A lower covering factor leads to slightly colder dust, and a higher fraction of UV flux escaping, which is likely the case in low-metallicity environments. We normalise this luminosity at the wavelength of H$\alpha$. We expect the ionizing component to be in a thinner disc than the older stars, and so we used a vertical scale height of 50\,pc, half that of the 100\,Myr stars, and similar to the scale height of the UV discs \citep{2012Combes}.

\subsubsection{Dust System}\label{sec:dust_model}

\begin{figure*}
\centering
\includegraphics[width=2\columnwidth]{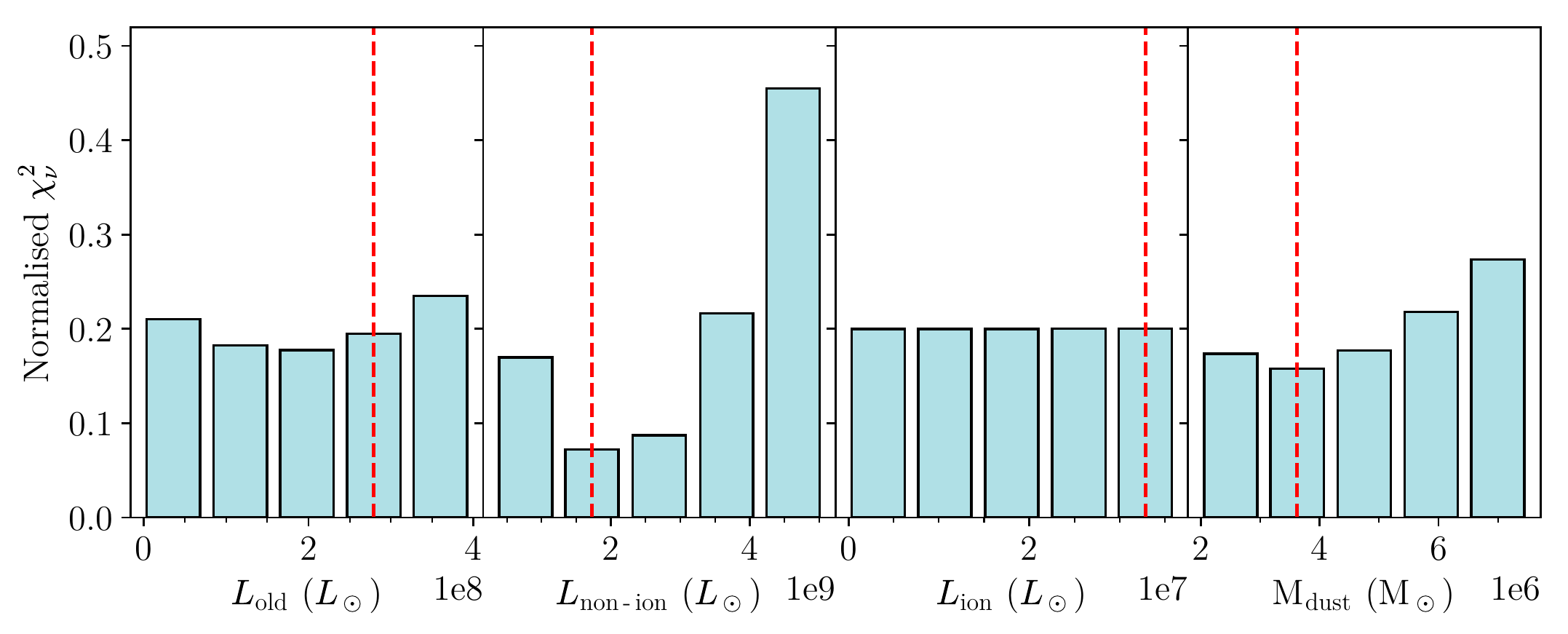}
\caption{Normalised $\chi_\nu^2$ distributions of the four free parameters in our model. From left to right, these are the old stellar luminosity, the non-ionizing young stars, the ionizing stars and the dust mass. The distribution is normalised such that the total sum of the  bars is equal to one, so that the four distributions are at approximately the same scale. The dashed red line shows the best fit parameter used in the high-resolution simulation.}
\label{fig:chi_sq}
\end{figure*}

We created a map of the dust mass as the input component for the dust geometry. For this, we made use of pixel-by-pixel {\sc magphys} fits. The dust model is more thoroughly described in \cite{2008DaCunha}, but as a brief overview, {\sc magphys} models Polycyclic Aromatic Hydrocarbons (PAHs) using a fixed template based on M17, which dominate at MIR wavelengths. The hot dust is modelled as a series of modified blackbodies with temperatures of 850, 250 and 130\,K. We use a {\sc magphys} library with extended priors \citep{2014Viaene}, meaning that the warm dust is modelled as a modified blackbodies (MBBs) with a fixed $\beta$ of 1.5, and can vary from 30 to 70\,K. The cold dust has a fixed $\beta$ of 2 and can vary from 10 to 30\,K. We use a vertical scale height of 100\,pc, the same as the young, non-ionizing stellar population, and similar to the 100\,\micron\ scale height found by \citep{2012Combes}. This is also consistent with model predictions of edge-on galaxies \citep[e.g.][]{1999Xilouris,2008Bianchi,2013DeGeyter}.

This model implicitly assumes a per-pixel local dust-energy balance. As we discuss in Sect. \ref{sec:dust_energy_balance}, this is not an acceptable assumption at scales <1500\,pc. However, as shown in \cite{2018Williams}, there is a very tight relationship between the dust masses obtained from MAGPHYS, and from a single-temperature MBB, with a median offset of 0.02\,dex and an RMS scatter of 0.10\,dex for the same data (their Fig. 9). As an additional check, we also performed this comparison on a pixel-by-pixel dust map fitted using our modified THEMIS fitting routine (Appendix \ref{app:themis_fitting}), and find a similar relationship. Thus, this choice of dust map will have a negligible impact on the simulation.

We use the THEMIS dust model \citep{2013Jones} to describe the dust in M33. This model consists of small and large amorphous hydrocarbons (sCM20 and lCM20), along with silicates \citep[aSilM5,][]{2014Kohler} to model the diffuse ISM of the MW. This model is primarily laboratory-based, and can naturally explain most of the features of the dust SED in the MW. However, in our initial testing we found that the default THEMIS parameterisation was insufficient to fit the dust SED of M33, particularly at sub-mm wavelengths, where M33 is known to have a sub-mm excess \citep{2016Hermelo,2018Relano}. We therefore modified THEMIS from its default parameters, which is described in more detail in Appendix \ref{app:themis_fitting}. The main results of this modification are to use a dust mix with fewer very small carbon grains, which we might expect in a low-metallicity environment where the dust grains have less shielding from the interstellar radiation field (ISRF). The fit also modifies the silicate-to-carbon ratio. In the MW, this is $\sim$10 but we find a mass ratio of $\sim$0.3, very similar to the ratios found in the LMC and SMC by \cite{2017Chastenet}. This would imply that either silicate grains are readily destroyed, or do not form in great numbers. This is unlikely, and so more likely is that the silicate grains are not emissive enough in the current THEMIS model, as inferred from more recent laboratory studies (Anthony Jones, priv. comm.). Given more emissive silicate grains, a smaller mass of carbon grains would be required to explain the flatter sub-mm slope, and this ratio would be closer to that of the MW. It is, therefore, not necessarily a much higher mass of carbon grains that are required, but simply a higher mass of more emissive dust grains. The ratio of small-to-large grains is very similar to the MW, however (0.4 in the MW, 0.3 in our fitting). As shown in Fig. \ref{fig:themis_sed_fit}, the parameters of the THEMIS dust model can be adjusted to fit well in the sub-mm range. The results of this fitting confirm the hypothesis of \cite{2016Hermelo}, who suggest different physical grain properties as the most plausable explanation for the observed sub-mm excess in M33. We use these recalculated abundances and size distributions in our {\sc skirt} model, but within the RT simulation keep this dust mix constant throughout the entire galaxy.

\section{Model Fitting}\label{sec:model_fit}

\begin{figure*}
\centering
\includegraphics[width=2\columnwidth]{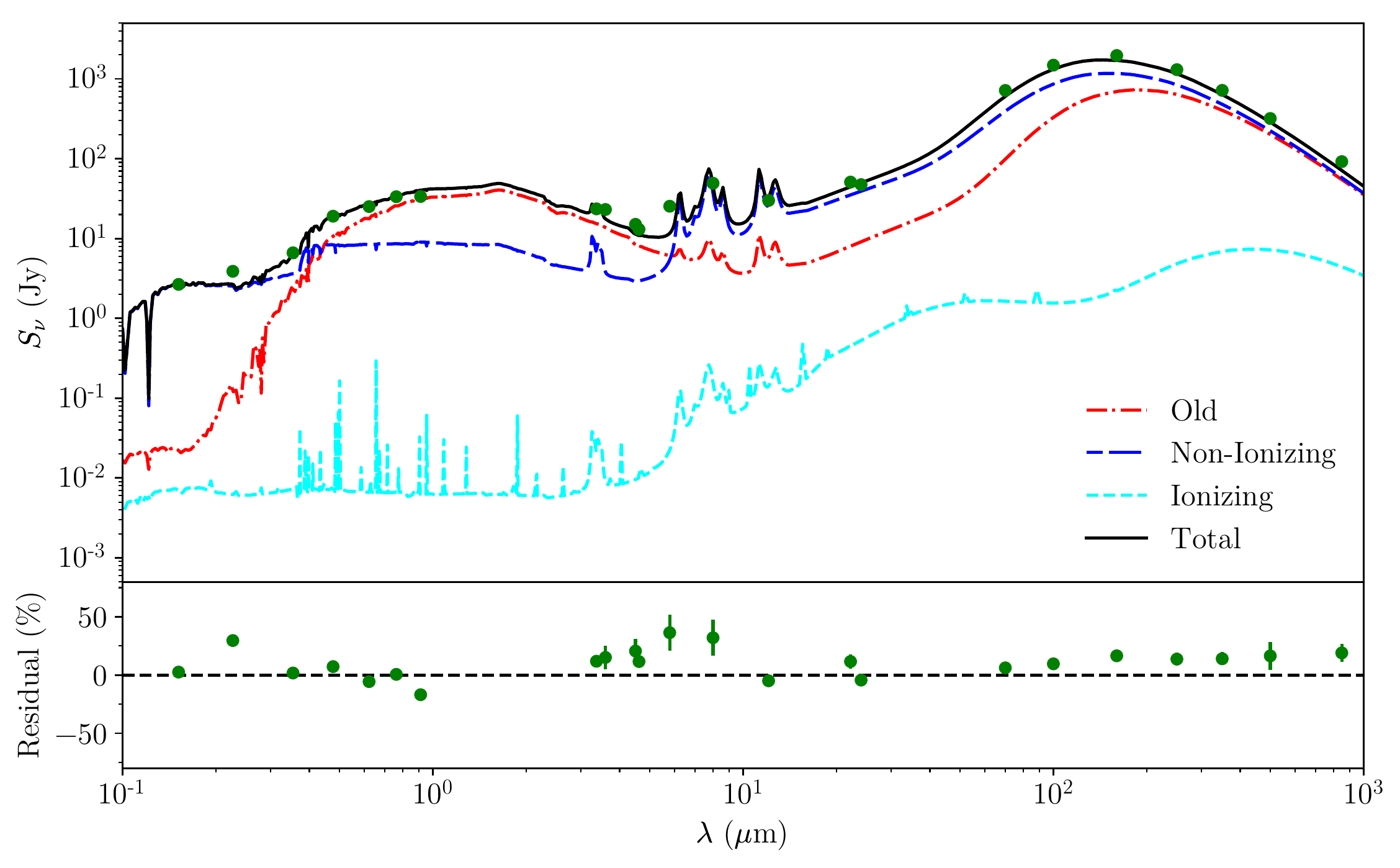}
\caption{\textit{Top:} High-resolution, best-fit RT simulation for M33. The red dot-dash line indicates the contribution from old stellar populations, the short-dash long-dash blue line young non-ionizing stars, light blue dashed line the ionizing population and the solid black line the total. Since the simulation considers the dust heating from the three stellar components simultaneously, this black line is not simply the sum of the three component lines. In this sense, the component decomposition should be taken as indicative only. \textit{Bottom:} residuals for this fit.}
\label{fig:skirt_sed}
\end{figure*}

To find the best fit model, we ran a series of simulations with a variety of luminosities exploring the parameter space around our initial guesses (see Table \ref{tab:skirt_parameters}). As these simulations are computationally expensive, and we can only explore the parameter space using a grid method, we fix all of the parameters apart from the various normalisations. For each of these parameters, we use 5 equally spaced values between our minima and maxima, for a total of 625 simulations.

We ran our simulations using a wavelength grid of 90 points, spaced for effective convolution with various filters and weighted depending on the importance of photons in that particular energy regime (Verstocken et al., in prep). We also use a small number of photons ($10^6$), to reduce the computational time for each model. Our dust grid is a binary tree dust grid (see \citealt{2014Saftly} for more details on this grid method), and cells are no longer split when their mass fraction is less than 10$^{-5}$. This means that cells are not equal in size, with smaller cells in areas of higher density. In total, each of these ``low-resolution'' simulations takes around 3 CPU hours, and contain around 150,000 dust cells. To determine the best fit model, we defined six wavelength regimes -- UV (GALEX), optical (SDSS), NIR (3.4 -- 4.6\,\micron), MIR (5.8 -- 24\,\micron), FIR (70 -- 250\,\micron), and submm (250 -- 850\,\micron). We calculated the reduced chi-squared, $\chi_\nu^2$, for each of these wavelength regimes, as there are an uneven number of points in each wavelength range. Our best fit is then the minimum of the sum of each of the $\chi_\nu^2$ values, including an extra 10\% error in each of the points to account for uncertainties in the modelling, which is often used in other studies \citep[e.g.][]{2009Noll}. Table \ref{tab:skirt_parameters} gives our best fit values for each free parameter. The likelihood of each model is given by $\mathcal{L} \propto \exp^{-\chi_\nu^2/2}$, and we sample from this distribution, quoting our errors as the 16$^{\rm th}$ and 84$^{\rm th}$ percentiles. When this error is smaller than the width of the bin, we instead quote the error as half the bin width. We find that given the low luminosity of the ionizing stars, we cannot well constrain this parameter, and it has a flat $\chi_\nu^2$ distribution across our parameter range, but our other parameters are reasonably well constrained. The normalised $\chi_\nu^2$ distributions of these free parameters are shown in Fig. \ref{fig:chi_sq}.

Having found a best-fit model, we then re-simulated this model using a higher resolution wavelength grid with 553 points, and $2 \times 10^7$ photons to produce images with more reliable filter convolution and higher S/N. The cell maximum mass fraction is decreased by a factor of 10, to 10$^{-5}$, which leads to around 1 million dust cells. These cell sizes vary from 37\,pc$^3$ to 1500\,pc$^3$, with an average size of 51\,pc$^3$. With a maximum optical depth in a cell of 0.47, and an average V-band optical depth of 0.01, we can be confident that this grid is well sampled, even in regions of high optical depth. We also break the simulation up into its various stellar components to see the relative contribution from each across our wavelength range. We use an instrument referred to as a {\sc FullInstrument} in {\sc skirt}, which separates the total recorded flux, at every pixel and at every wavelength, in contributions due to direct stellar flux, scattered stellar flux, direct dust flux, and scattered dust flux, i.e. 
\begin{equation}
S_\lambda^{\text{tot}} = S_\lambda^{\star, {\rm dir}} + S_\lambda^{\star, {\rm sca}} + S_\lambda^{{\rm dust, dir}} + S_\lambda^{{\rm dust, sca}}.
\end{equation}
This instrument also calculates $S_\lambda^{\text{tra}}$, i.e. the flux that would be obtained if the galaxy were completely dust-free. These simulations take around 600 CPU hours each.

\section{Results and Discussion}\label{sec:results}

\subsection{Global SED}

The high-resolution, best-fit SED can be seen in Fig. \ref{fig:skirt_sed}. We also repeated the simulation for each stellar component individually, to illustrate the contribution from each of these components. As discussed in Sect. \ref{sec:dust_heating}, this decomposition will not take into account the fact that the dust is simultaneously heated by each component. Thus, the sum of these lines will not be equal to the overall SED. We calculated the residuals by convolving the overall SED with the respective filter response for each waveband to produce a model flux, and then
\begin{equation}
    {\rm residual} = \frac{{\rm observation} - {\rm model}}{{\rm observation}}.
\end{equation}
In this sense, a negative residual means the model overestimates the observed data, and vice-versa. In the UV, the emission is dominated by light from the young stellar populations. In the optical and NIR, the emission is dominated by the old stellar populations. The dust emission is, in general, dominated by heating from the young stars, but is formed of a complex interplay of heating from the stellar components -- a warmer component from heating due to the young stellar populations and a colder component from heating due to the older stars. The dust heating from the ionizing stars forms two distinct bumps, one from warmer dust heated from within the molecular clouds, and a cooler component from the emission of the more diffuse dust in the ISM surrounding these birth clouds.

\begin{table}
 \caption{SFRs for M33, calculated using a variety of SFR tracers. In each case, we give the model SFR (SFR$_{\rm model}$), the SFR as calculated from the data (SFR$_{\rm obs}$) and references for the SFR prescription used. SFRs are in M$_\odot$\,yr$^{-1}$.}
 \label{tab:sfr_comparison}
 \centering
 \begin{tabular}{cccc}
  \hline
  \hline
  Band(s) & SFR$_{\rm model}$ & SFR$_{\rm obs}$ & Reference\\
  \hline				
  24\,\micron & 0.11 & $0.10\pm0.01$ & a\\
  70\,\micron & 0.14 & $0.15\pm0.02$ & b\\
  FUV+24\,\micron & $0.23^{+0.04}_{-0.02}$ & $0.25^{+0.10}_{-0.07}$ & c\\
  TIR & 0.23 & $0.17\pm0.06^1$ & d, e\\
  \hline
  \multicolumn{4}{p{.8\columnwidth}}{(a) \cite{2009Rieke}, (b) \cite{2010Calzetti}, (c) \cite{2008Leroy}, (d) \cite{2011Hao}, (e) \cite{2011Murphy}. $^1$Including an error of 30\% to estimate uncertainty in IMF, dust attenuation; single temperature modified blackbody.}
 \end{tabular}
\end{table}

We find a median absolute deviation (MAD) across all wavebands of 12\%. We find that the NUV point is underestimated in the model (with a residual value of 30\%). The RT model underestimating the NUV point is common across similar studies \citep[see, e.g.][]{2014DeLooze,2016Mosenkov,2017aViaene}, and is likely caused by a UV attenuation bump that is too strong (see Sect. \ref{sect:dust_attenuation}). We also find that many of the MIR points are underestimated. The MIR points are dominated by aromatic features, and so producing an adequate fit in this wavelength range is strongly dependent on the properties of the small carbon grains. Increasing the weighting to these points can produce a better fit at these wavelength ranges, but a much poorer fit to the UV/optical points. Given the complex nature of this wavelength range, and our particular interest in the local dust-energy balance of M33, we prefer the current fit. Finally, the longer wavelengths are underestimated, potentially indicating a dust mass that is too low, or an incorrect dust emissivity. However, an increase in dust mass leads to increased dust attenuation and a poorer fit to the short wavelength points. Considering the uncertainty on the power-law slope for the small carbon grains (4.26$\pm$0.13), and the fact that this has a large effect on the dust emissivity, the emissivity could well be underestimated. Given the fact the UV/optical and FIR/sub-mm points are given equal weight, this is the preferable fit. Due to recent work on the sub-mm excess \citep{2016Hermelo,2018Relano}, we explore the 450 and 850\micron\ wavelengths in more detail in Sect. \ref{sect:submm_excess}.

\begin{figure*}
\centering
\includegraphics[width=2\columnwidth]{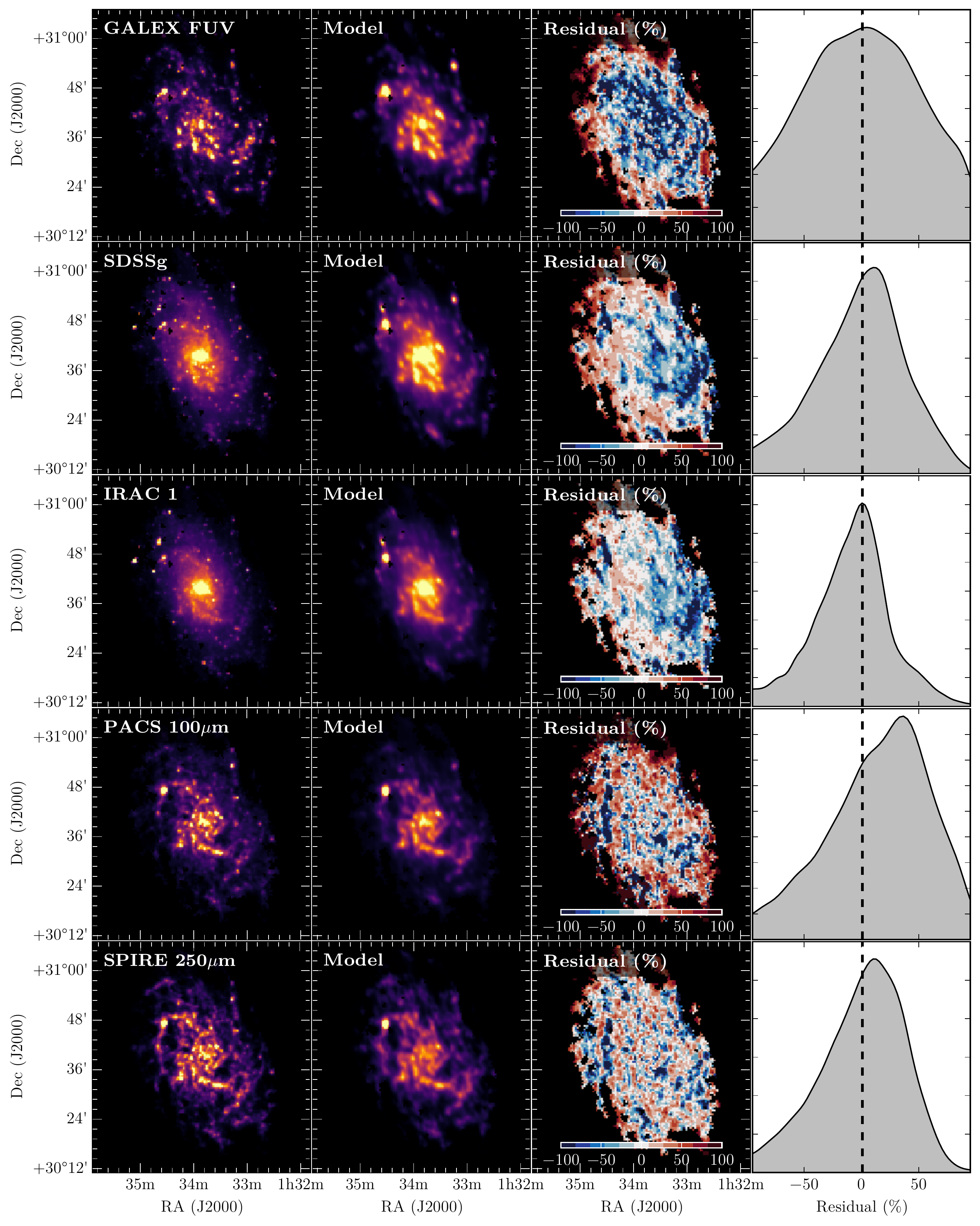}
\caption{Residual plots of {\it first row:} GALEX FUV, {\it second row:} SDSS g-band, {\it third row:} IRAC 3.6\micron, {\it fourth row:} PACS 100\micron, {\it fifth row:} SPIRE 250\micron\ wavebands. In each case, we show {\it first column:} the observed image, {\it second column:} the model image from our RT simulation, {\it third column:} the residuals, and {\it fourth column:} a KDE plot of these residuals.}
\label{fig:residuals}
\end{figure*}

We calculated the SFR that this model produces using a variety of single- and multi-band SFR tracers (24\,\micron, 70\,\micron, a combination of FUV+24\,\micron, and TIR luminosity), and compared these to the values calculated from the data. The results of this can be seen in Table \ref{tab:sfr_comparison}. We see a good correspondence between the modelled and observed SFRs in M33. The TIR SFRs marginally agree within error, but we note a difference in the way these are calculated -- for SKIRT, we integrate the emission from 3-1100 micron to calculate a TIR flux. In the case of the observed data, we fit a single-temperature modified blackbody (MBB) to the cold dust continuum emission, which will have negligible contributions at shorter wavelengths. As the longer wavelength regime is more affected by dust heating from older stellar populations, with a higher fraction of dust heating at shorter wavelengths from the young stellar populations (see Sect. \ref{sec:dust_heating}), our model TIR luminosity is likely more representative of the TIR luminosity. We also highlight the importance of including the unattenuated starlight here -- compared to the monochromatic 24\,\micron\ and 70\,\micron\ calculated SFR, the FUV+24\,\micron\ SFR is nearly a factor of 2 higher. This is also true for our observed SFRs. The calculated SFR is consistently lower (by a factor of 2-3) than those calculated by \cite{2009Verley}. The reason for this is twofold -- firstly, they use SFR prescriptions similar to that of \cite{1998Kennicutt}. Our updated SFR measurements are generally around a factor of two lower (see \citealt{2012KennicuttEvans}, their Table 1). Secondly, in truncating and masking the disc, we remove a significant amount of flux in the outer disc. The values given in Table \ref{tab:sfr_comparison}, therefore, should be treated as a consistency check between pixels considered in the simulation and observations, and not as a measure of the true SFR of M33.

\begin{figure*}
\centering
\includegraphics[width=2\columnwidth]{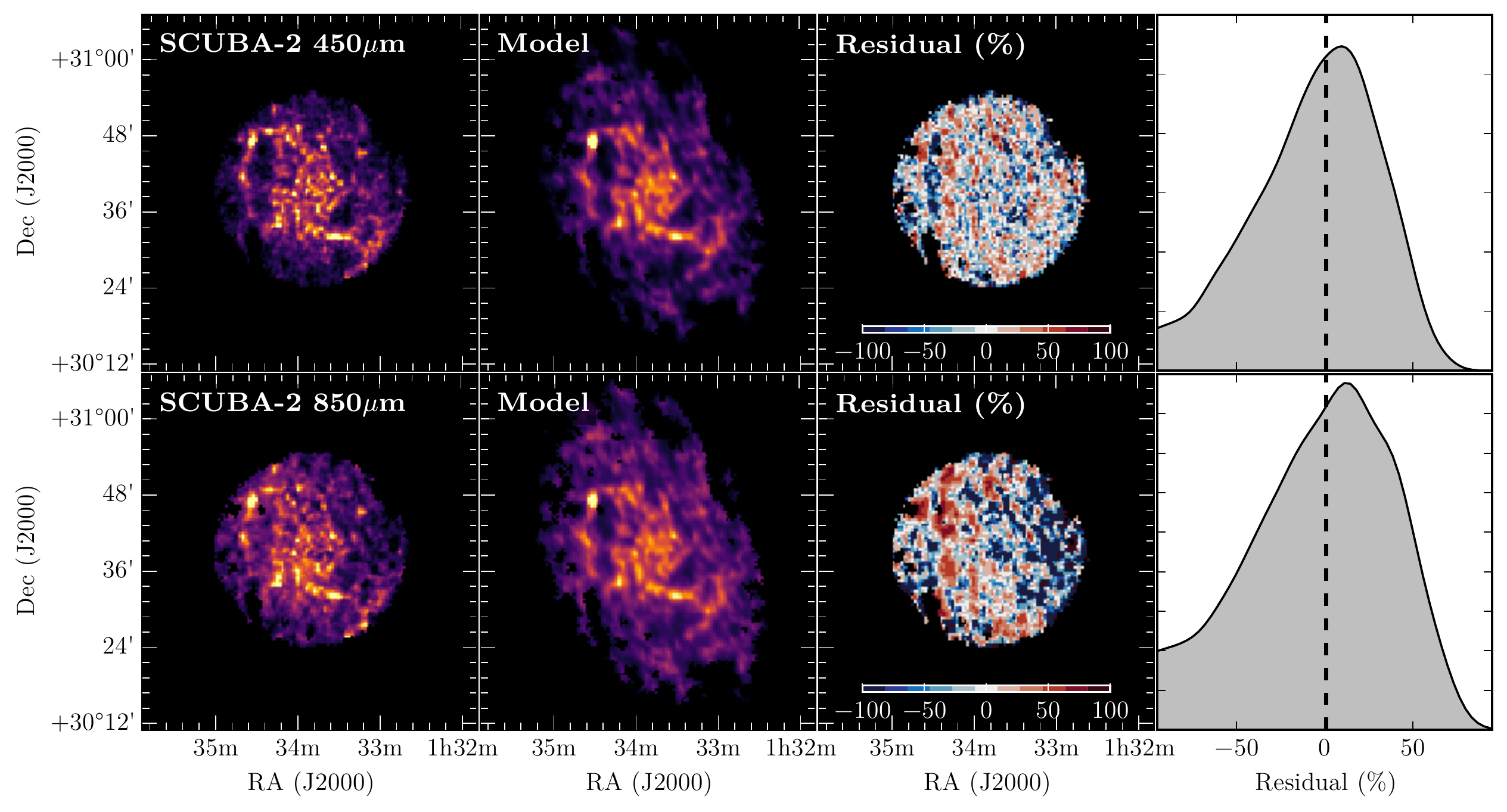}
\caption{Residual plots of {\it top:} SCUBA-2 450\micron\ and, {\it bottom:} SCUBA-2 850\micron\ wavebands. We show {\it first column:} the observed image, {\it second column:} the model image from our RT simulation, {\it third column:} the residuals, and {\it fourth column:} a KDE plot of these residuals.}
\label{fig:residuals_submm}
\end{figure*}

Alternatively, the SFR can also be calculated directly from the SED templates inputted into {\sc SKIRT} for the young stellar populations, as these are scaled from a known normalisation factor. For the non-ionizing stars (i.e. the SFR over 100\,Myr), this gives a value of 0.15 M$_\odot$\,yr$^{-1}$, similar to the single-band SFR prescriptions in Table \ref{tab:sfr_comparison}. For the ionizing stars (the SFR over the last 10\,Myr), this produces an SFR of 0.58 M$_\odot$\,yr$^{-1}$, but given that the ionizing stellar luminosity is not well constrained, this is unlikely to be a good measure of the true SFR.

From this model, we can calculate the fraction of the stellar radiation absorbed by dust, $\mathcal{F}_{\rm abs}$, which is given by
\begin{equation}
    \mathcal{F}_{\rm abs} = \frac{L_{\rm dust}}{L_{\rm dust}+L_{\rm stars}}.
\end{equation}
For the DustPedia galaxies, \cite{2018Bianchi} find this value to be 19\% on average (25\% if only considering late-type galaxies). For M33, we find this value to be 21\%, in agreement with the findings of this earlier study.

We find a dust mass of $(3.6\pm0.6) \times 10^6\,{\rm M}_\odot$. This is very comparable to \cite{2018Williams}, with a dust mass calculated from {\sc magphys} of $4 \times 10^6\,{\rm M}_\odot$, for a similarly good fit to this wavelength regime (see Fig. \ref{fig:themis_sed_fit} and Fig. 2 of \citealt{2018Williams}). However, \cite{2016Hermelo} find a significantly higher dust mass, of around $1.7 \times 10^7\,{\rm M}_\odot$ (albeit with a large uncertainty). However, as discussed in their work, this leads to a much lower gas-to-dust ratio (GDR) than expected in this low-metallicity environment, and so this dust mass estimate is likely too high, potentially due to grain properties or the dust attenuation assumed in their models. Assuming that the GDR scales with metallicity as $Z^{-1}$ \citep[e.g.][]{2007Draine,2011Leroy}, we would expect a value of the GDR between 200-450 \citep{2016Hermelo}. Using a total gas mass of $1.7\times10^9$\,M$_\odot$ \citep{2010Gratier,2014Druard}, we get a GDR of 470, significantly higher than \cite{2016Hermelo}, but in agreement with the radial profiles of \citet[][their Fig. 10]{2018Relano}.

\subsection{A Resolved Comparison of M33}

{\sc skirt} also produces a data cube which provides a 2D view of the galaxy at each wavelength in the wavelength grid. This means that we can compare the model on a spatial scale at any given wavelength. To this end, we produced residual images at a number of wavelengths. We note that due to images going through rotation and projection within the {\sc skirt} routine, comparing these images directly may be an unfair comparison. This is discussed in more detail in Appendix \ref{app:rotate_project_data}, but the effect of this on a moderately inclined galaxy like M33 is minor, and so we opted to compare directly to the original images. Given the resolution of our input geometries, we first spectrally convolve these datacubes with the relevant filter response, before spatially convolving with the point-spread function (PSF) of that waveband and regridding to pixels of 25\,arcsec\, (using {\sc montage}) to make these images comparable. We also mask any pixels not considered in our simulation.

We find that across the 23 wavebands that form the high-resolution dataset for our simulation, we have a MAD of 33\%, higher than the deviation seen in our global fluxes. Plots of the residuals at five wavelengths (FUV, SDSSg band, 3.6\micron, 100\,\micron, and 250\,\micron) are shown in Fig. \ref{fig:residuals}. In general, our residuals are centred around 0 and most of the values lie within $\pm$50\% of the observed values. We see strong structure in many of our residuals, with the model often overestimating in the spiral arms and underestimating in the more diffuse ISM. In the regimes where we are observing mainly starlight, this is likely due to our temporal and spatial resolution. Whilst we assume 3 discrete, average ages, the actual star-formation history is much more complex, with stars of similar ages clumping together \citep[e.g.][]{2015Lewis}, and these variations are on scales smaller than we are able to model in our simulation. At the wavelengths when we are dominated by dust emission, these spatial variations might indicate a variation in dust grain properties. Previous work has shown that there can be significant variation in the dust properties across a galaxy (e.g. \citealt{2012Smith} in M31, \citealt{2014Tabatabaei} in M33). \cite{2018Relano} also suggest regional variations in dust properties to better explain the sub-mm excess. This work, however, has assumed an average dust grain property and mix throughout the whole of M33. We do not believe the features present in the residuals are an artefact of the use of {\sc magphys}, as these features are also present in the study of \cite{2014DeLooze}, where the geometries are defined in a completely independent way to our analysis. There is also noise inherent both in the observations and the simulation, which makes a resolved comparison difficult. However, despite the simplicity of the model, the simulations well resemble the observations.

\subsubsection{The sub-mm excess}\label{sect:submm_excess}

Given the sub-mm excess present in M33 \citep[e.g.][i.e. that the observed fluxes are higher than the model]{2016Hermelo,2018Relano}, we have also produced residuals for the model at 450\micron, and 850\micron\ wavelengths. These are compared to our SCUBA-2 images, and can be seen in Fig. \ref{fig:residuals_submm}. Unlike \cite{2016Hermelo}, we find no significant sub-mm excess in our model (any higher than the excess we have at all long-wavelength points), and we also find no clear radial dependence on our residuals (consistent with those seen in any other wavelength regime), unlike that of \cite{2018Relano}. However, we note that in our earlier THEMIS fitting we modify the dust grain properties {\it specifically} to fit the sub-mm excess by removing many of the small carbon grains, and so the fact that we do not see this excess is not surprising.

\begin{figure}
\centering
\resizebox{\hsize}{!}{\includegraphics{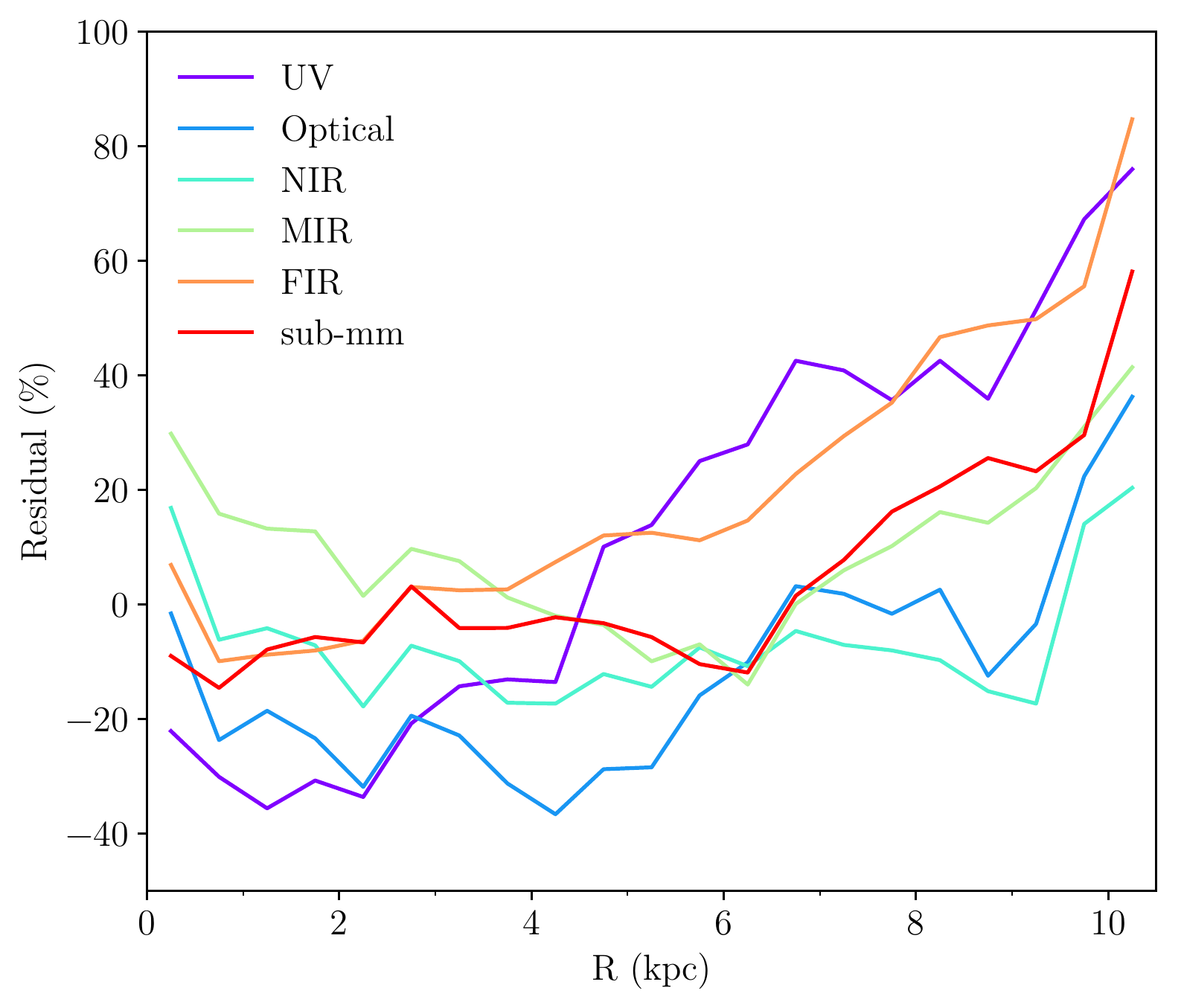}}
\caption{Median residual with galactocentric radius. Each differently coloured line indicates a wavelength regime as defined in Sect. \ref{sec:model_fit}.}
\label{fig:radial_residuals}
\end{figure}

\subsubsection{Radial variation of the residuals}\label{sect:radial_variation}

\begin{figure}
\centering
\resizebox{\hsize}{!}{\includegraphics{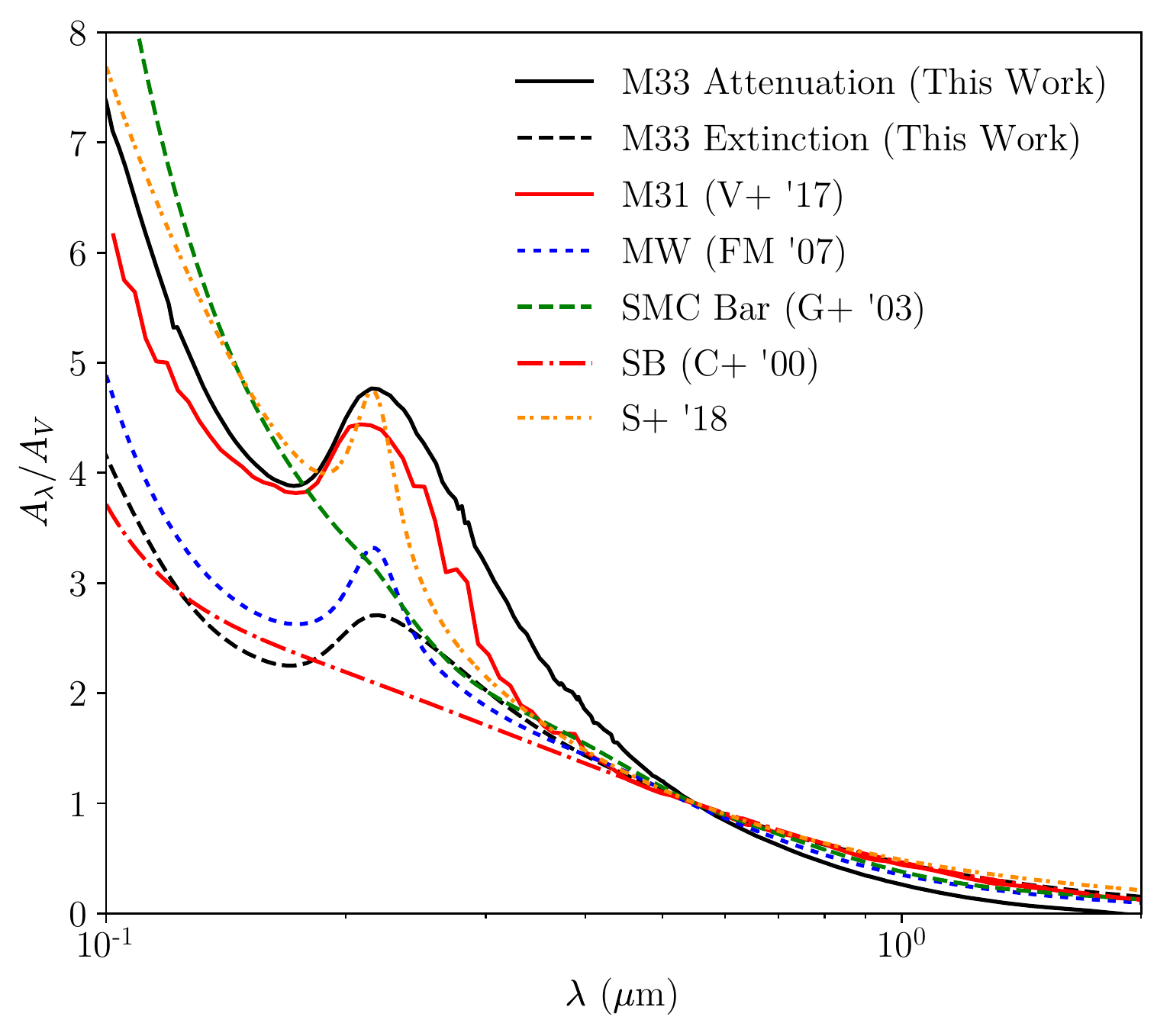}}
\caption{Dust attenuation for our RT model (solid black line) and underlying extinction curve (dashed black line), compared to extinction curves for the M31 (solid red line, \citealt{2017aViaene}), the MW (short-dashed blue line; \citealt{2007FitzpatrickMassa}), the SMC bar region (long-dashed green line; \citealt{2003Gordon}), the \citet{2000Calzetti} law for starburst galaxies (dot-dash red line), and the derived attenuation curve of \citet{2018Salim} for the stellar mass of M33 (short dash-dot orange line). All of these curves are normalised at V-band.}
\label{fig:dust_attenuation}
\end{figure}

\begin{figure*}
\centering
\includegraphics[width=2\columnwidth]{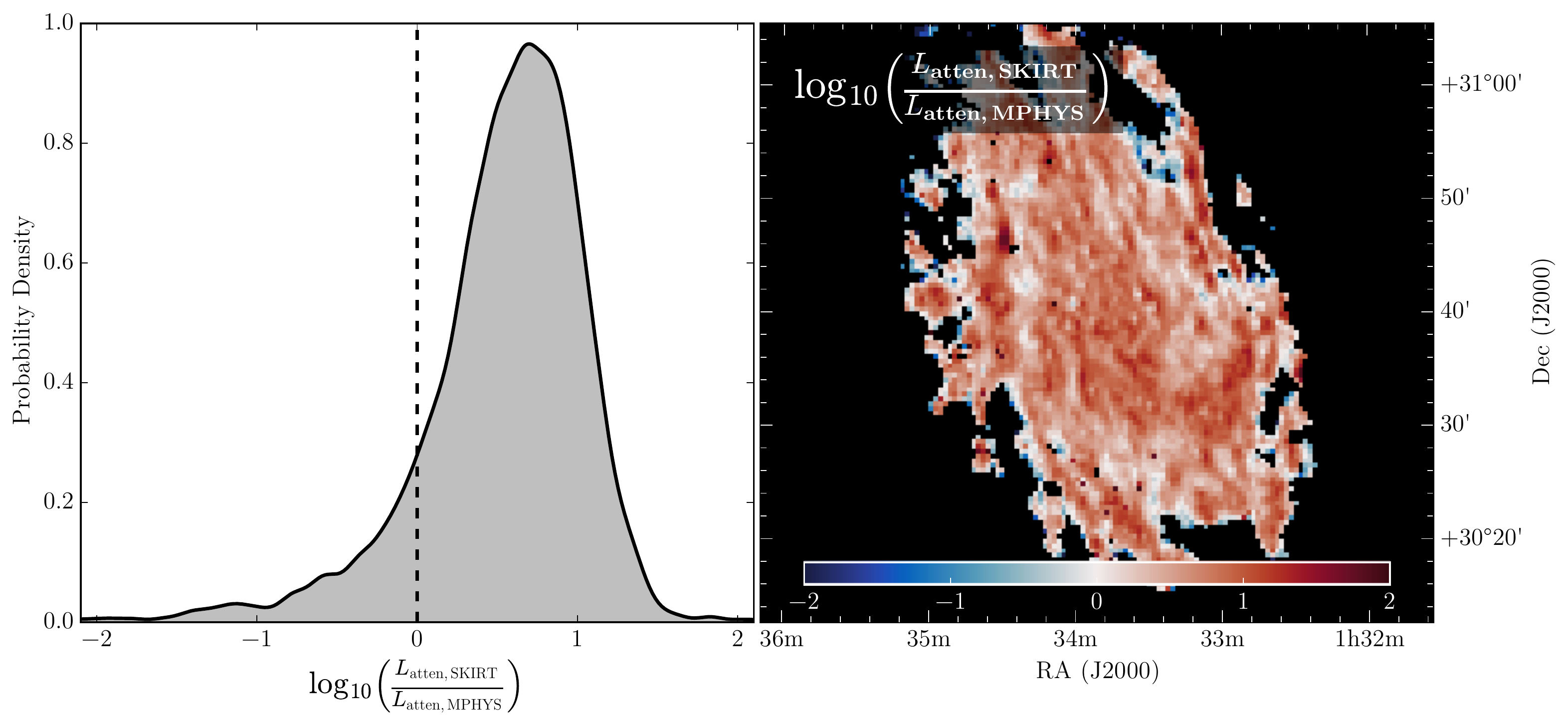}
\caption{Comparison of {\sc skirt} and {\sc magphys} attenuated luminosities. {\it Left:} KDE plot of the ratio of the {\sc skirt} attenuated luminosity to that of {\sc magphys}. The black dashed line indicates where these two values are equal. {\it Right:} this same ratio, but instead plotted positionally.}
\label{fig:attenuation_skirt_mphys}
\end{figure*}

M33 has been shown to have a warped disc, both in the optical \citep[e.g.][]{1980SandageHumphreys}, and in its H{\sc i} 21cm line \citep[e.g.][]{1997CorbelliSchneider}. If the inclination and position angle assumed in the deprojection of our model inputs has some radial variation, we would expect that to be seen as some radial dependence in the residuals. To quantify this, we calculate the median residual for each wavelength regime as defined in Sect. \ref{sec:model_fit} for bins of 0.5\,kpc width in deprojected galactocentric radius (assuming a position angle of 22.5\degr, and inclination of 56\degr). The results of this are shown in Fig. \ref{fig:radial_residuals}. We see that in general, the model tends to underestimate at low galactocentric radius, and increasingly overestimate with increasing galactocentric radius. Given that the trends seen are broadly similar between each wavelength regime, we therefore conclude that M33 is similarly warped across all wavelengths considered in this study.

\subsection{Dust Attenuation}\label{sect:dust_attenuation}

\subsubsection{Global Attenuation}

By using the {\sc FullInstrument} instrument class type in {\sc skirt}, a view of the system with no dust attenuation is produced as a result of the simulation. From this, we can directly calculate the dust attenuation at each wavelength, as
\begin{equation}
A_\lambda = 2.5 \log_{10}\left(\frac{S_\lambda^{\rm unatten}}{S_\lambda^{\rm atten}}\right),
\end{equation}
where $S_\lambda^{\rm unatten}$ and $S_\lambda^{\rm atten}$ are the total unattenuated and attenuated fluxes from the simulation at a given wavelength, respectively. One important caveat for the attenuation is that the MAPPINGS III SED does not truly provide a transparent view of the system with no dust, as the dust is built-in to this SED type. We make an estimate of the effect this will have on the flux by comparing a ``transparent'' MAPPINGS SED (i.e., a covering factor of 0) with the SED produced by our adopted covering factor of 0.1. This makes an average difference of 10\% to the fluxes, which given the much lower luminosity of the ionizing stellar populations will make a negligible difference to our results. Along with calculating an attenuation curve, we also extract the intrinsic dust extinction curve, to see the effects of geometry and scattering on this attenuation curve. The results of this are shown in Fig. \ref{fig:dust_attenuation}, and are compared to several literature extinction curves.

Visually, the underlying extinction curve lies somewhere between the MW \citep{2007FitzpatrickMassa} and starburst galaxies \citep{2000Calzetti}. We find a reasonably strong UV bump, which is broader than that of the MW. This is due to the adopted dust grain properties in the THEMIS model, which are discussed in more detail in \cite{2013Jones}. This figure also highlights the effect of the treatment of geometry and dust scattering in shaping the difference between attenuation and extinction curves. These properties have previously been shown to have an important role in shaping attenuation curves \citep[e.g.][]{2000Granato,2000WittGordon,2001BaesDejonghe,2007Panuzzo,2017bViaene}. Our M33 dust attenuation curve appears much more similar to the SMC bar region of \cite{2003Gordon}, albeit with a strong UV bump. This is somewhat counter-intuitive, as our fitting technique alters the size distribution of the very small carbon grains, which are responsible for this bump. However, the larger of these small grains ($1\,{\rm nm} < r < 20\,{\rm nm}$) also contribute to this bump \citep[][their Table 1]{2013Jones}, so removing the smallest grains will not necessarily eliminate this UV bump. The attenuation curve is also very similar to the attenuation curve calculated by \cite{2016Viaene} for M31, derived in a similar way. This is somewhat surprising, given the very different intrinsic properties and geometry between M31 and M33. Given the stellar mass of M33 (3-6$\times 10^9$ M$_\odot$, \citealt{2003Corbelli}), the average dust attenuation curve for this stellar mass from the work of \cite{2018Salim} is very similar to that obtained in our simulations, although our UV bump is wider. An analysis of the nuclear region of M33 \citep{1999Gordon} finds evidence of strong attenuation, along with a strong 2175\AA\, bump. Our RT simulation shows that this may be the case across the whole of M33. However, we note that as the NUV flux is underestimated in our simulation, the strength of this NUV bump may be overestimated.

\subsubsection{Comparison to SED Modelling}

We can also compare, positionally, the amount of dust attenuation in our RT simulation to more traditional SED fitting. For our comparison, we take the pixel-by-pixel {\sc magphys} fitting from \cite{2018Williams}. {\sc magphys} uses the dust attenuation model of \cite{2000CharlotFall}, and we refer readers to that work for details of the model. Essentially, this model assumes two populations of stars -- one in their birth clouds, and others that have drifted away from these birth clouds. The light from both of these populations is subject to attenuation from dust in the ISM, and the stars in their birth clouds are additionally attenuated by the dusty clouds they reside within. This attenuation has a power-law type dependence on the wavelength, and the V-band optical depth is one of the parameters {\sc magphys} fits, as well as the fraction of attenuation by dust in the ISM compared to birth clouds. The \cite{2000CharlotFall} model, or variations of it, are typically used in panchromatic SED fitting tools. As {\sc magphys} employs a dust-energy balance, the modelled dust luminosity is by definition the attenuated luminosity. For {\sc skirt}, in terms of its {\sc FullInstrument} output,
the attenuated luminosity is
\begin{equation} \label{eq:attenuated_luminosity}
    L^{\text{atten}} = 4\pi\,D^2 \int \left( S_\lambda^{\text{tra}} - S_\lambda^{\star, \text{dir}} - S_\lambda^{\star, \text{sca}} \right) d\lambda
\end{equation}
Given that this does not include the flux directly from the dust, this is not simply the transparent flux minus the total flux in the simulation. We calculate the ratio of the {\sc skirt} to {\sc magphys} attenuated luminosities, and show this in Fig. \ref{fig:attenuation_skirt_mphys}. We find a median offset of 0.56 dex for the {\sc skirt} luminosity compared to the {\sc magphys} attenuated luminosity, and a clear positional dependence in this offset, with much higher values for {\sc skirt} in the spiral arms, and regions of more intense star formation, as compared to {\sc magphys}. The reason for this may be two-fold -- firstly, the pixel-by-pixel {\sc magphys} fitting uses pixels of 100\,pc$^2$, where the local dust-energy balance may not hold (i.e. the amount of dust luminosity and attenuated luminosity may not be the same). With simulations of a galaxy, \cite{2018Smith} find acceptable fits to the V-band attenuation on scales of 0.2-25\,kpc in $\sim$99\% of pixels modelled with {\sc magphys}, so this is unlikely to explain the large discrepancy between these two attenuated luminosities. Secondly, the geometry can play an important role in affecting dust attenuation -- given the positional dependence on the discrepancies between {\sc magphys} and {\sc skirt}, this is more likely to be the case.

\subsubsection{Face-On Optical Depth}

\begin{figure}
\centering
\resizebox{\hsize}{!}{\includegraphics{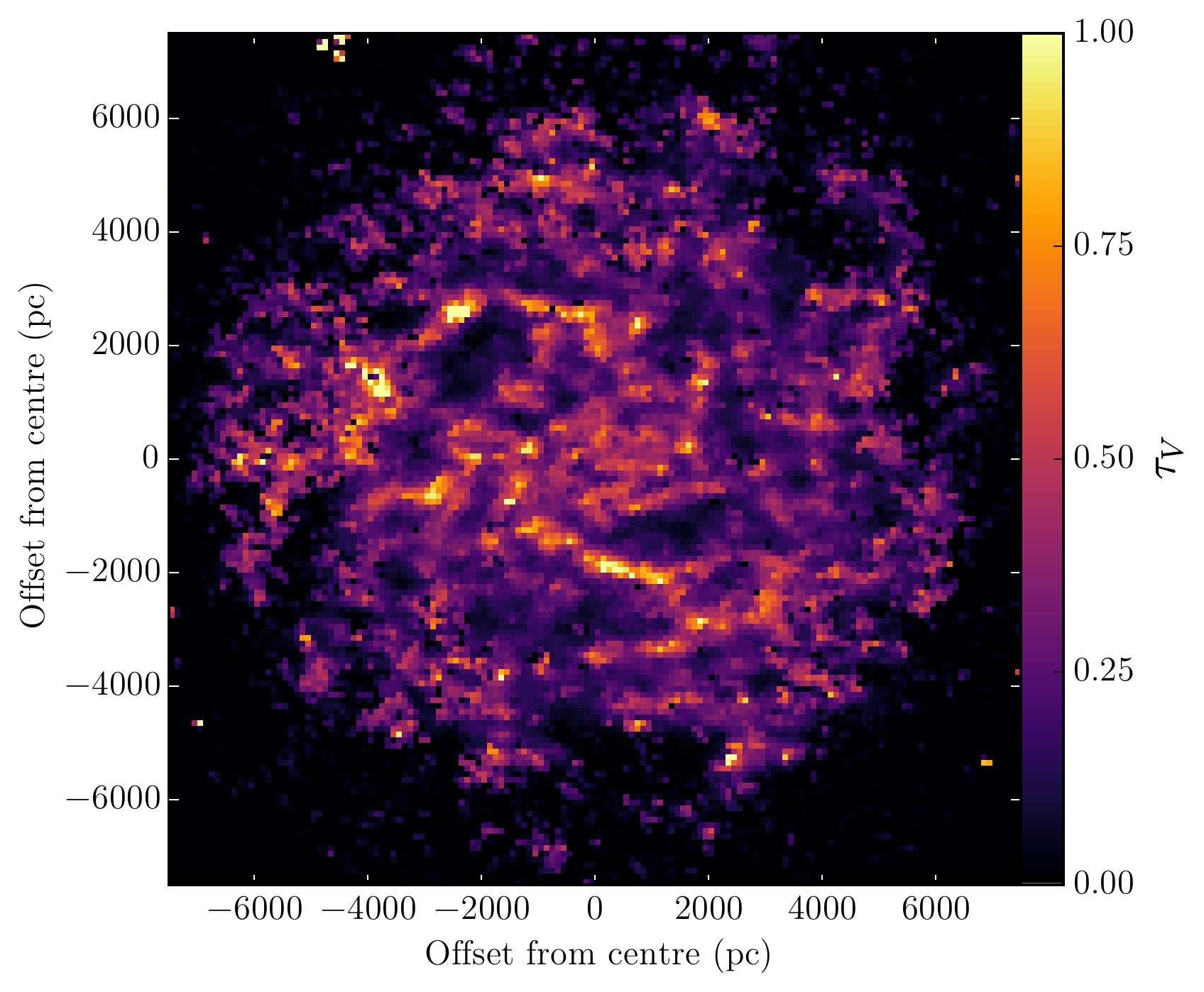}}
\caption{Face-on optical depth in V-band, derived from the RT simulation.}
\label{fig:optical_depth}
\end{figure}

The optical depth of a galaxy is an important parameter to measure, as it quantifies the amount of stellar light that can be viewed directly without being obscured by dust. The question of whether galaxies are optically thin when viewed face-on is an outstanding problem, with some studies claiming the disc is nearly transparent \citep[e.g.][]{1999Xilouris}, and some claiming that galaxies tend to be optically thick \citep[e.g.][]{1997Trewhella}. Several works have attempted to answer this question through RT modelling of edge-on galaxies \citep[e.g.]{2014DeGeyter,2018Mosenkov}, but given degeneracies between the dust scale-length and the face-on optical depth, a reliable estimate of $\tau_V$ has been difficult to ascertain.

The optical depth is simply given by the dust column density integrated along the path length of a photon, and multiplied by the extinction coefficient, i.e.:
\begin{equation}
    \tau_V = \kappa_V \int_0^\infty \rho_{\text{d}}(s)\, ds,
\end{equation}
where $\kappa_V$ for our dust mixture is calculated in the simulation to be 4625\,m$^2$\,kg$^{-1}$. Taking a deprojected column density map, this can then be trivially converted into a map of the optical depth, and we show this in Fig. \ref{fig:optical_depth}. This map shows that the optical depth is highest in the spiral arms, and peaks in areas of active star formation. This peak can reach values >1, and thus these regions are optically thick. However, across the spiral arms the average optical depth is $\sim$0.3, in the interarm regions are $\sim$0.1, and the average V-band optical depth across the whole galaxy is $\sim$0.2. There is a gentle radial decline with galactocentric radius, from $\sim$0.5 in the centre to $\sim$0.2 at a radius of 5\,kpc. This is well in agreement with \cite{2009Verley}, who find a decrease in $A_V$ with increasing galactocentric radius. We therefore conclude that at scales of 100pc, M33 is generally optically thin across its disc.

\subsection{Dust Heating Mechanisms}\label{sec:dust_heating}

\begin{figure}
\centering
\resizebox{\hsize}{!}{\includegraphics{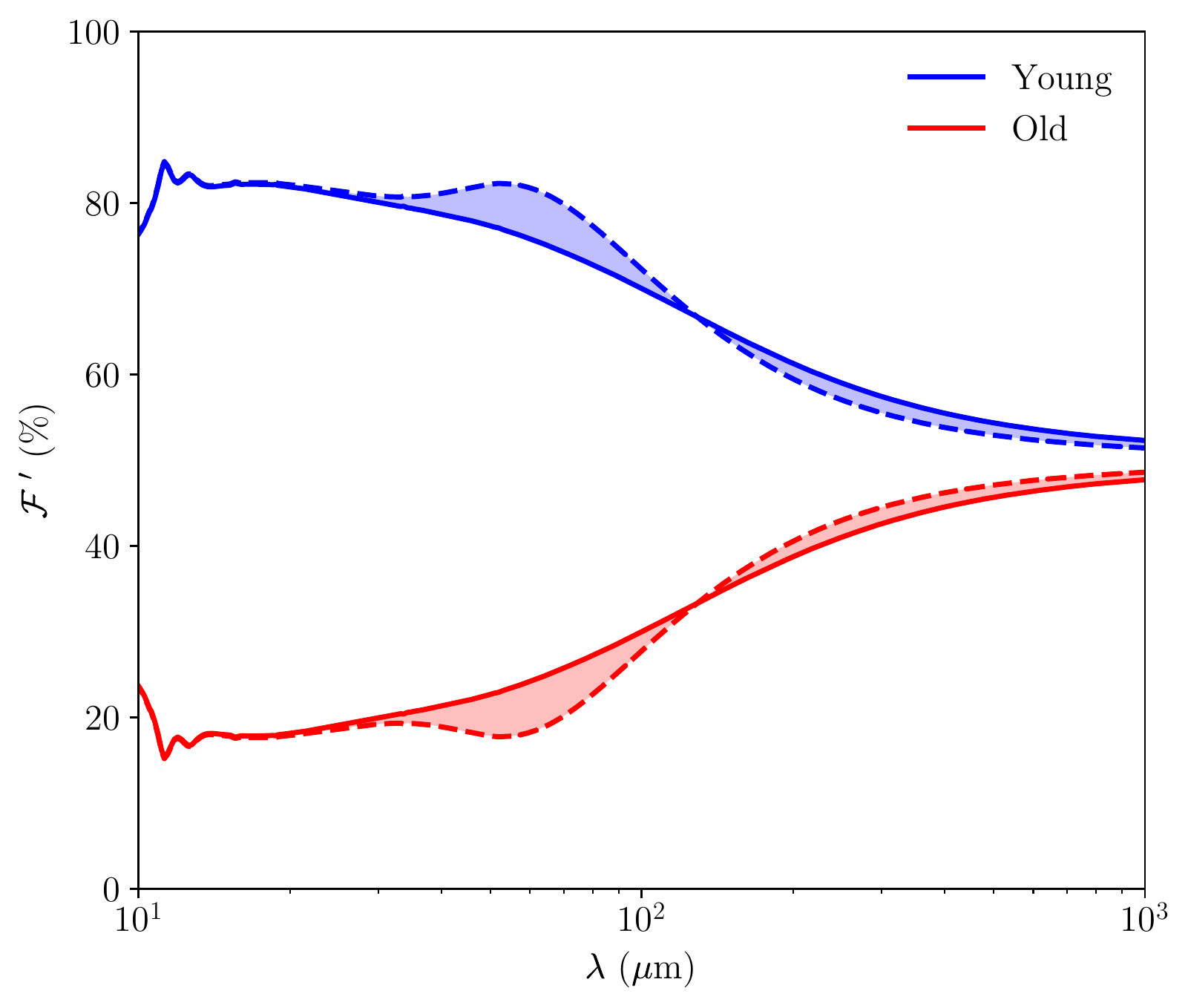}}
\caption{Dust heating fraction with wavelength from the young (blue) and old (red) stellar populations. In each case, the solid line corresponds to Eqs. \ref{eq:f_prime_young} and \ref{eq:f_prime_old}, and the dashed line to Eqs. \ref{eq:f_prime_young_uncorr} and \ref{eq:f_prime_old_uncorr}.}
\label{fig:dust_heating_wavelength}
\end{figure}

\begin{figure*}
\centering
\includegraphics[width=2\columnwidth]{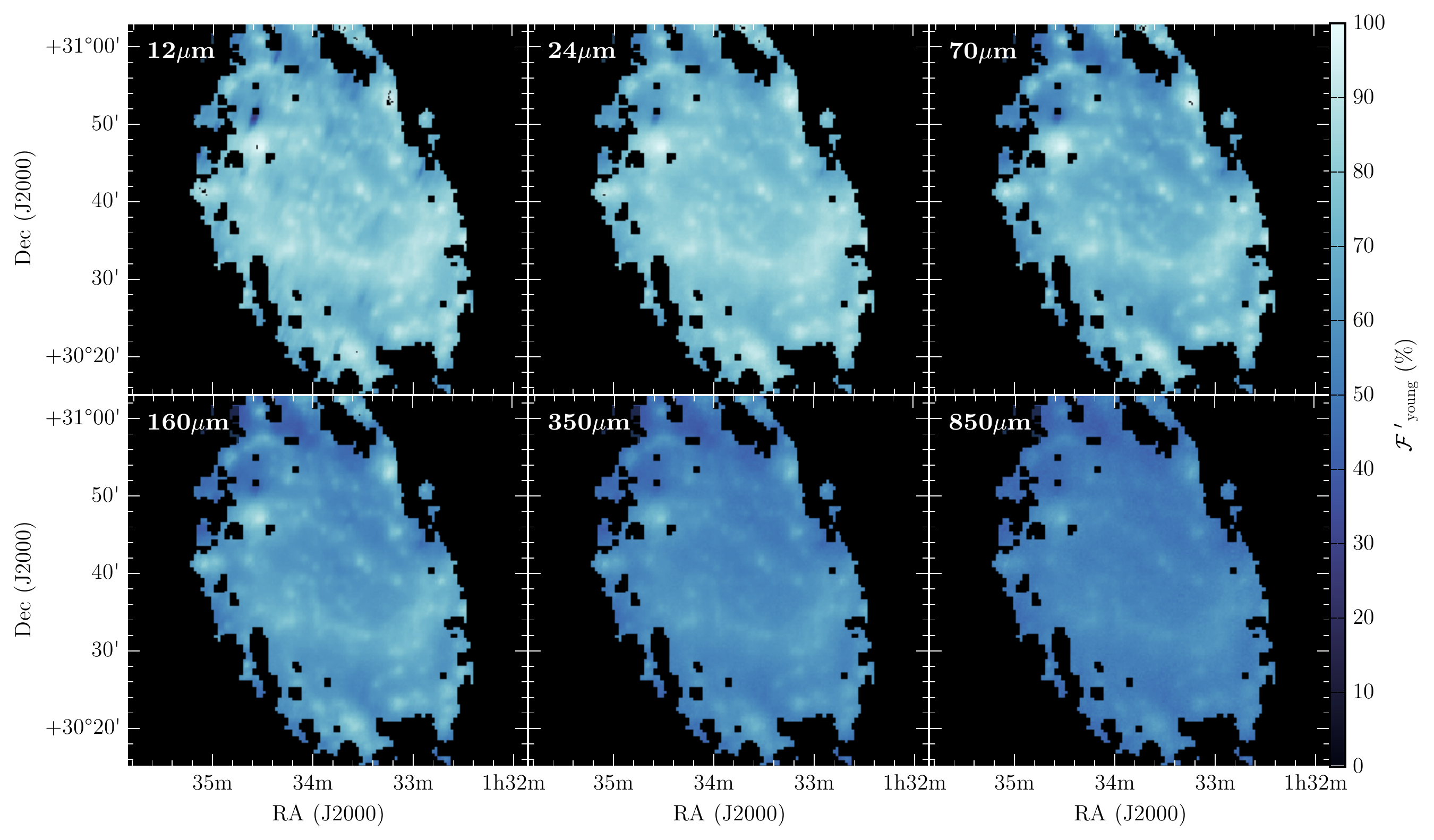}
\caption{Dust heating fraction by young stellar populations ($F^\prime_{\lambda, {\rm young}}$) for a selection of wavebands. From left to right, \textit{top row:} WISE 12\,\micron, MIPS 24\,\micron, and MIPS 70\,\micron. \textit{Bottom row:} PACS 160\,\micron, SPIRE 250\,\micron, and SCUBA-2 850\,\micron.}
\label{fig:dust_heating_position}
\end{figure*}

From the {\sc skirt} model, we can calculate the fraction of dust heating that comes from the young versus the old stars ($\mathcal{F}_{\rm young}$ and $\mathcal{F}_{\rm old}$, respectively), as simply the ratio of the total dust luminosity from the young stars divided by the total dust luminosity of the simulation. We find $\mathcal{F}_{\rm young}$ to be 72\%, similar to the 63\% found by \cite{2014DeLooze} for M51, but significantly higher than the 9\% found by \cite{2016Viaene} for M31. Globally, the dust heating of M33 is driven mainly by the young stellar populations.

We next turn to the contributions to the dust heating by the various stellar populations on a resolved level. Whilst $\mathcal{F}_{\rm young} + \mathcal{F}_{\rm old} = 1$ on a global SED scale, due to the intertwined nature of the radiation fields and the non-locality in wavelength of the dust heating, separating these quantities can only be approximated. We used the approximation of \cite{2014DeLooze}, where
\begin{equation}\label{eq:f_prime_young}
    \mathcal{F}^\prime_{\lambda, {\rm young}} = \frac{1}{2} \frac{ S_{\lambda, {\rm young}} + \left(S_{\lambda, {\rm total} } - S_{\lambda, {\rm old} }  \right)}{ S_{\lambda, {\rm total}} },
\end{equation}
and 
\begin{equation}\label{eq:f_prime_old}
    \mathcal{F}^\prime_{\lambda, {\rm old}} = \frac{1}{2} \frac{ S_{\lambda, {\rm old}} + \left(S_{\lambda, {\rm total} } - S_{\lambda, {\rm young} }  \right)}{ S_{\lambda, {\rm total}} }.
\end{equation}
Due to the non-local nature of the dust heating, $S_{\lambda, {\rm total}}$ is not simply $S_{\lambda, {\rm old}} + S_{\lambda, {\rm young} }$. We also included the na\"{i}ve case where we do not include this non-locality, i.e.
\begin{equation}\label{eq:f_prime_young_uncorr}
    \mathcal{F}_{\lambda, {\rm young}} = \frac{ S_{\lambda, {\rm young}} }{ S_{\lambda, {\rm young}} + S_{\lambda, {\rm old}} },
\end{equation}
and
\begin{equation}\label{eq:f_prime_old_uncorr}
    \mathcal{F}_{\lambda, {\rm old}} = \frac{ S_{\lambda, {\rm old}} }{ S_{\lambda, {\rm young}} + S_{\lambda, {\rm old}} }.
\end{equation}
The results of this are shown in Fig. \ref{fig:dust_heating_wavelength}. At all wavelengths, the dust heating is driven mainly by the young stellar populations, with a decreasing contribution towards longer wavelengths. It appears that contributions to the dust heating from the old stellar populations peak at colder dust temperatures, as they are heating the colder, more diffuse dust in the ISM \citep[e.g.][]{2008Bianchi,2015Natale,2015Bendo}.

It is also possible to investigate the fractional contribution to the dust heating from the young stellar populations on a resolved basis. Using Eq. \ref{eq:f_prime_young}, we calculate $F^\prime_{\lambda, {\rm young}}$ across a number of wavebands, and the results of this can be seen in Fig. \ref{fig:dust_heating_position}. It can be seen that at shorter wavelengths, there is a higher contribution to the dust heating from the young stellar populations in the spiral arms of M33, but this discrepancy decreases with increasing wavelength, to an almost uniform distribution at 850\micron. This is similar to that observed by \cite{2017aViaene}, where the rings of M31 are clearly visible at shorter wavelengths.

\subsection{Local Dust-Energy Balance}\label{sec:dust_energy_balance}

\begin{figure}
\centering
\resizebox{\hsize}{!}{\includegraphics{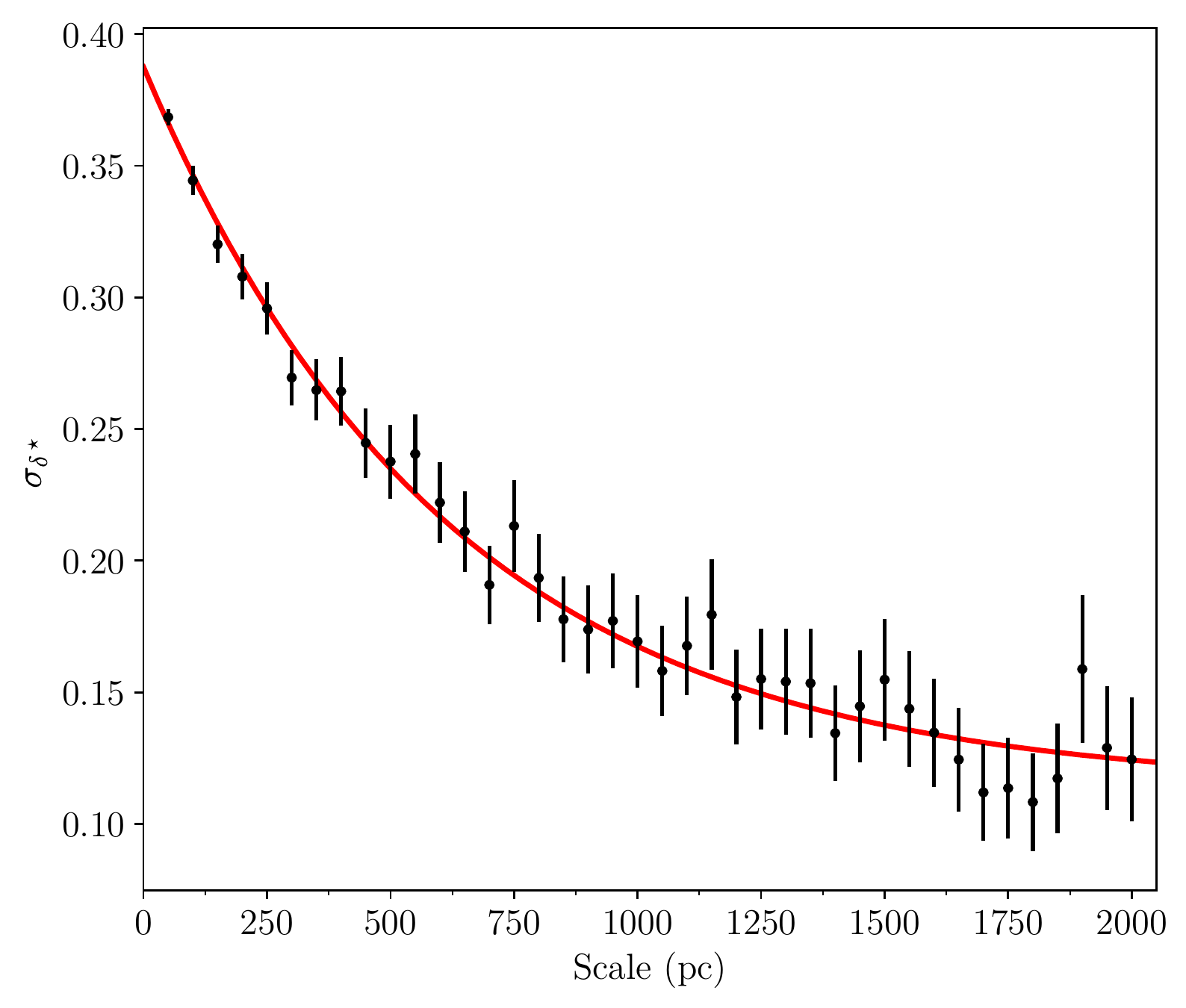}}
\caption{$\sigma_{\delta^\star}$ for a variety of spatial scales. Errors are plotted assuming normal errors. The values flatten at a scale of around 1500pc, indicating this is the scale at which the local dust-energy balance holds true. The red line shows a fitted exponential, intended to guide the eye.}
\label{fig:dust_energy_balance}
\end{figure}

The scales at which the local dust-energy balance holds is vital for diagnosing the suitability of resolved measurements. In suitably small regions where more dust is heated by starlight originating from stars in neighbouring pixels than in the pixel being considered, traditional SED fitting tools may not recover a reliable value. We investigated the spatial scale at which the local dust-energy balance in our RT simulation becomes an acceptable assumption. This also gives an estimate of the average distance a photon travels within a galaxy. To this end, we define a ``stellar luminosity excess,''
\begin{equation}
    \delta^\star = \frac{L^{\rm atten} - L^{\rm dust}}{L^{\rm atten}},
\end{equation}
where $L^{\rm atten}$ is the total stellar luminosity attenuated (Eq. \ref{eq:attenuated_luminosity}), and $L^{\rm dust}$ is the luminosity emitted by the dust. In terms of the {\sc skirt} {\sc FullInstrument} output, this is
\begin{equation}\label{eq:dust_luminosity}
    L^{{\rm dust}} = 4\pi\,D^2 \int \left( S_\lambda^{{\rm dust, dir}} - S_\lambda^{{\rm dust, sca}} \right) d\lambda
\end{equation}
the integral of the direct flux from the dust. A value of 0 for $\delta^\star$ means the local dust-energy balance holds in that particular pixel, and increasingly positive (negative) values indicate more (less) flux attenuated than emitted by the dust in that pixel. Globally, the dust-energy balance should hold and therefore the mean of this distribution should be 0. We calculate this parameter for every 3D pixel in our data cube, and calculate the spread in these pixels, $\sigma_{\delta^\star}$, as the 84$^{\rm th}$ percentile minus the 16$^{\rm th}$ percentile. At the scale where the local dust-energy balance is a suitable assumption, $\sigma_{\delta^\star}$ should ideally be equal to 0. However, due to deviations between the model and observations, along with noise in the RT simulation, this is unlikely to be the case, so the point at which increasing the spatial scale causes no significant decrease in $\sigma_{\delta^\star}$ is the point at which we assume the local dust-energy balance takes hold. To calculate these parameters for a variety of spatial scales, we regrid the simulation output to a number of scales, rather than re-run the simulation many times. The results of this procedure for a variety of spatial scales is shown in Fig. \ref{fig:dust_energy_balance}. From this, we can see that the local dust-energy balance is a suitable assumption at scales greater than $\sim$1500\,pc. This is in agreement with simulations, which show that the local dust-energy balance holds true at scales greater than around 1\,kpc \citep{2018Smith}, as well as observational comparisons of SFR tracers \citep{2015Boquien}.

\section{Conclusions}\label{sec:conclusions}

In this work, we have presented a high-resolution (100\,pc) RT simulation of nearby spiral galaxy M33. Our simulation is self-consistent, includes the absorption and scattering effects of dust, and is performed in 3D. Our inputs for this simulation are based on images produced from a multi-wavelength dataset, in order to describe the young and older stellar populations in the galaxy. We also include a dust system, with the geometry informed by pixel-by-pixel SED fitting, and the dust properties from a global fit.

We find that we can well reproduce the SED of M33 to within a median deviation of 12\%. This SED is dominated across almost all wavelengths by the young stellar populations (direct emission at shorter wavelengths, and dust heating at longer wavelengths), and in the optical is characterised by strong dust attenuation, with a very strong and broad UV bump. We find that we can well reproduce the SFRs given by our observed data, as well as the total dust mass. However, we find discrepancies at a resolved level, with many wavebands showing strong features in the residuals. We argue that these are due to limitations in our simple model of this galaxy -- the stellar and dust properties are not homogeneous across the disc of M33, and appear to be strikingly different in the spiral arms versus the diffuse ISM, and not due to our choices of input geometries. We are able to fit the sub-mm excess detected in previous works with a modified THEMIS dust model, showing that the excess is related to a difference in the grain properties of the dust, as suggested by \cite{2016Hermelo}.

At a resolution of 100\,pc, the galaxy is mostly transparent in the V-band, except in areas of high star-formation. This means that we should reliably be able to calculate the stellar properties in galaxies at least to these scales. We also find that the dust is heated almost solely by the young stellar populations, and so the TIR luminosity should be a reasonable tracer for star formation in this galaxy.

Finally, we estimate that the local dust-energy balance does not hold below scales of around 1500pc. This means that tools that employ this balance (e.g. {\sc magphys}, {\sc cigale}) are likely to be unsuitable at these high resolutions.

Despite the simple nature of this RT model, we find that we can broadly reproduce the characteristics of M33. Even given its simplicity, this type of RT modelling allows insights into the sub-kpc properties of galaxies that traditional SED fitting does not, and allows us to probe the complex interplay of starlight and dust in galaxies self-consistently at these small spatial scales.

\section*{Acknowledgements}

The authors thank the anonymous reviewer, whose comments have certainly improved the manuscript. The authors also thank Peter Camps for valuable technical advice, along with all of the participants at the recent {\sc skirt} meeting for comments and discussions. I.D.L. gratefully acknowledges the supports of the Research Foundation -- Flanders (FWO). M.W.L.S acknowledges funding from the UK Science and Technology Facilities  Council consolidated grant ST/K000926/1. M.R. acknowledges support by the research projects AYA2014-53506-P and AYA2017-84897P from the Spanish Ministerio de Econom\'{i}a y Competitividad. This research made use of {\sc montage} (\url{http://montage.ipac.caltech.edu/ }), which is funded by the National Science Foundation under Grant Number ACI-1440620, and was previously funded by the National Aeronautics and Space Administration's Earth Science Technology Office, Computation Technologies Project, under Cooperative Agreement Number NCC5-626 between NASA and the California Institute of Technology. This research has made use of Astropy, a community-developed core Python package for Astronomy (\url{http://www.astropy.org/}; \citealt{2013Astropy,2018Astropy}). This research has made use of NumPy (\url{http://www.numpy.org/}; \citealt{2011vanderWalt}), SciPy (\url{http://www.scipy.org/}), and MatPlotLib (\url{http://matplotlib.org/}; \citealt{2007Hunter}). This research made use of APLpy, an open-source plotting package for Python (\url{https://aplpy.github.io/}; \citealt{2012Robitaille}).




\bibliographystyle{mnras}
\bibliography{bibliography} 



\appendix

\section{Modifying the THEMIS Dust Model}\label{app:themis_fitting}

\begin{figure*}
\centering
\includegraphics[width=2\columnwidth]{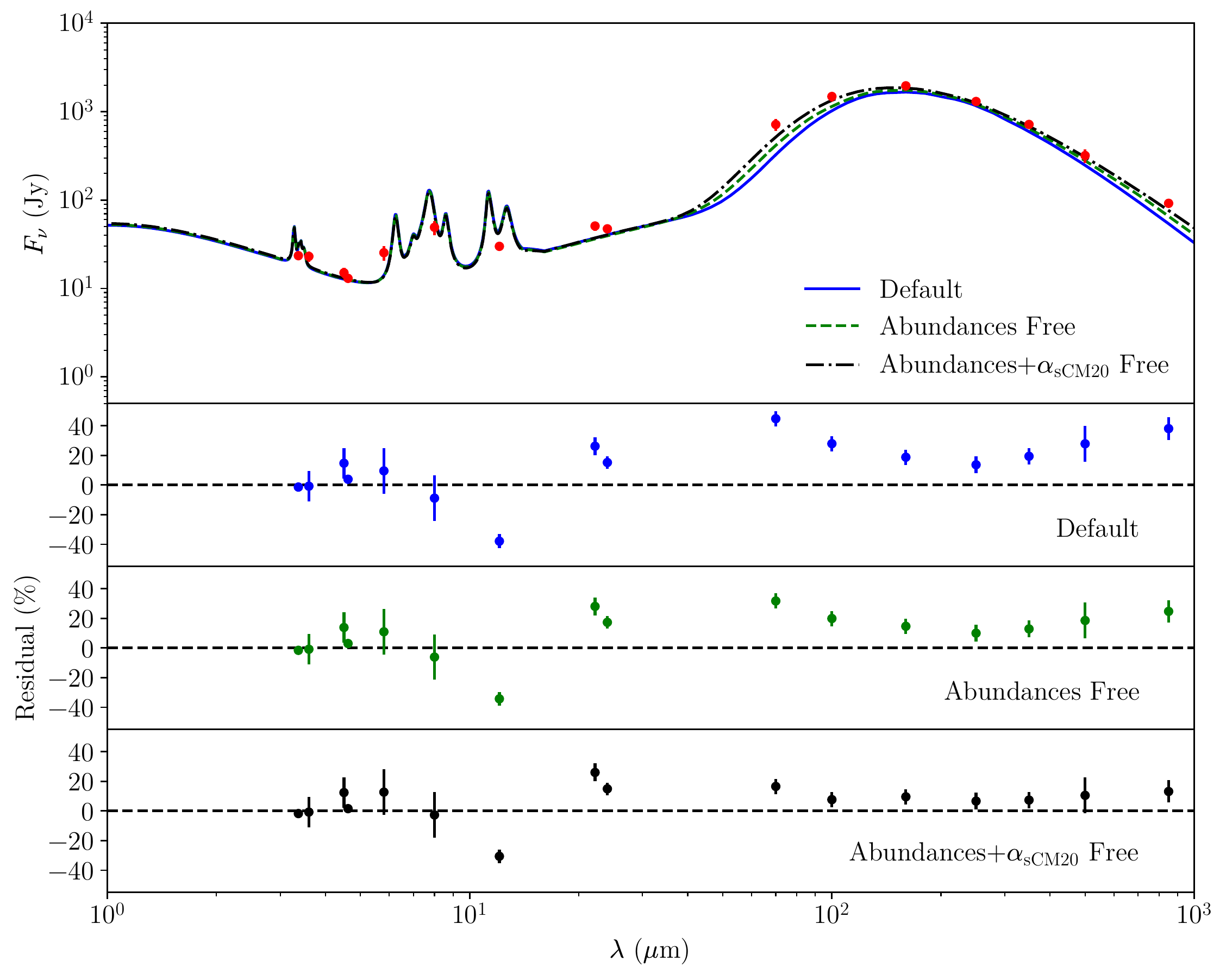}
\caption{\textit{Top:} Various model THEMIS fits to the dust SED of M33. The solid blue line indicates the default THEMIS parameters, the dashed green line where we allow mass abundances to vary, and the black dot-dash line where we additionally vary the size distribution of small carbon grains. The residuals for each of these fits are given in the subsequent panels.}
\label{fig:themis_sed_fit}
\end{figure*}

\begin{figure*}
\centering
\includegraphics[width=2\columnwidth]{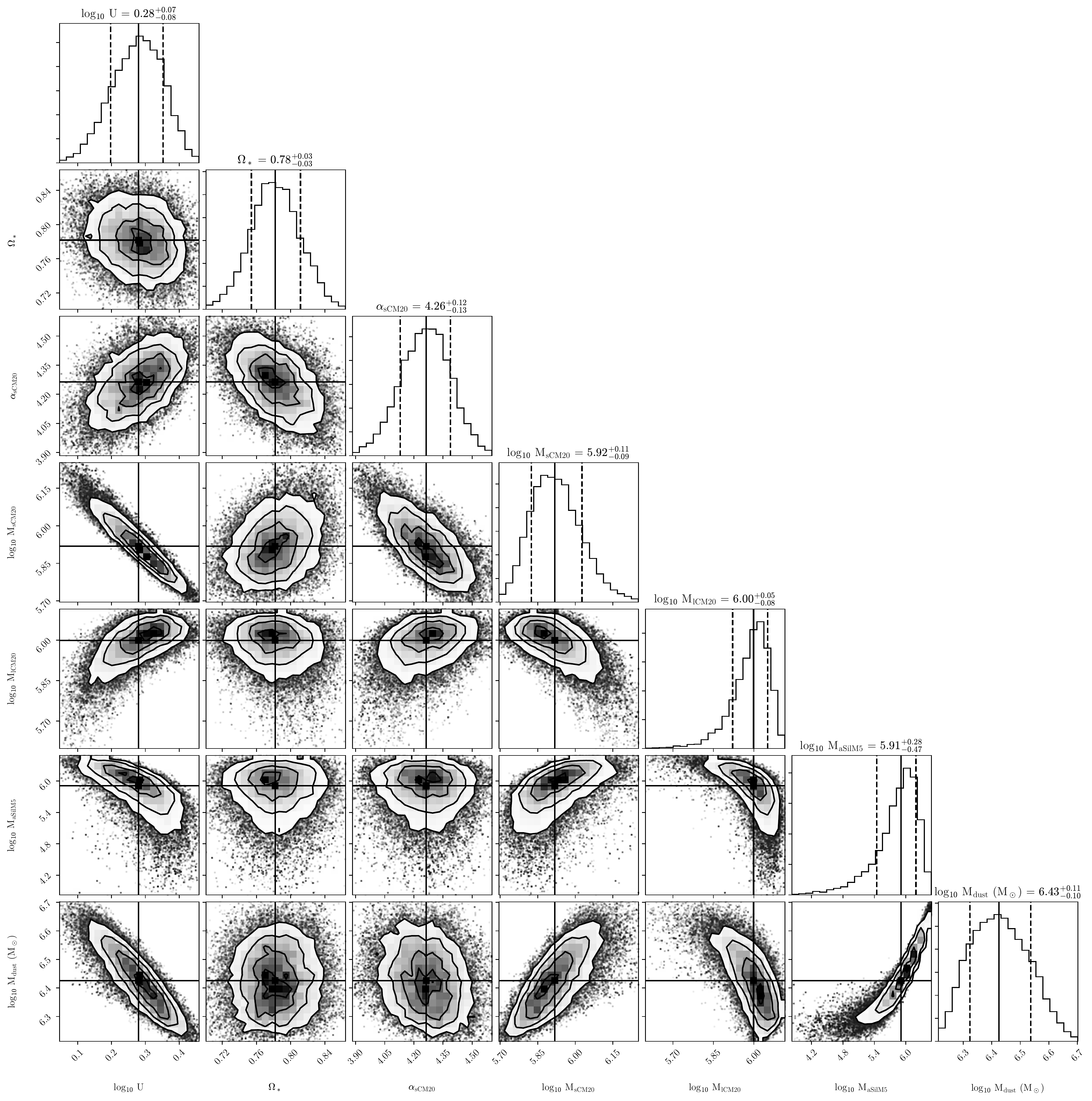}
\caption{Corner plot for the THEMIS dust fit with free dust grain abundances and small grain size distribution. From left to right, the panels show the ISRF strength U, the stellar scaling factor (with respect to the 3.6\,\micron\, flux), the small grain power-law slope, the mass for the small carbon grains, large carbon grains and silicates, and finally the total dust mass. The solid black line in each histogram shows the median value, with the dashed lines showing the 16$^{\rm th}$ and 84$^{\rm th}$ percentiles.}
\label{fig:themis_corner}
\end{figure*}

In order to better fit the THEMIS model to our data, we fit a dust emission model to the points longward of 3.4\,\micron\ using a similar method to that of \cite{2017Chastenet}. The components of this fit are the small carbon grains (sCM20), large carbon grains (lCM20), and silicates, consisting of olivines and pyroxines, tied together as aSilM5. Due to a non-negligible contribution from stars at NIR wavelengths, we also include a blackbody at 5000\,K, to approximate this contribution.

Our initial fit kept the mass ratios of these various components fixed at the values calculated for the diffuse dust of the MW, and so there are only three free parameters in the model -- the strength of the ISRF, a scaling factor for the stellar contribution and the overall dust mass. We generate a grid of ISRFs from $10^{-1} \leq U \leq 10^{3.5}$ (with 1 being the value for the local neighbourhood), equally spaced in steps of 0.01 in log space. The SEDs for this are generated using {\sc DustEM} \citep{2011Compiegne}. We then fitted these three free parameters using an MCMC framework using {\sc emcee}\footnote{\url{http://dfm.io/emcee/current/}}. We use 500 walkers each taking 500 steps using the first half of these steps as ``burn-in'', and our initial guesses for the ISRF is that of the MW, the stellar scaling factor the 3.6\,\micron\ point, and the dust mass by the 250\,\micron\ point. We account for correlated uncertainties between bands, and use the filter responses for each waveband to calculate the flux as seen by that particular instrument. The fit and residuals for this can be seen in Fig. \ref{fig:themis_sed_fit}, and we find that the default THEMIS parameters consistently underestimate the bulk of the cold dust points.

Next, we performed a fit where we allowed the abundances of the amorphous hydrocarbons and silicates to vary (although we lock the abundances of the two silicate populations together). This increases our number of free parameters to 5, where compared to the total dust mass we now have the individual masses of the small and large carbon grains, and the silicates. This fit is also shown in Fig. \ref{fig:themis_sed_fit}, and we find that while it performs slightly better than the default parameters, the fit is still poor across the FIR/sub-mm range.

Finally, we allowed variation in the small grain size distribution. The size distribution of small amorphous hydrocarbons is given by a power-law, partly defined by $\mathrm{d}n/\mathrm{d}a \propto a^{-\alpha_\mathrm{sCM20}}$, and we allow this value of $\alpha_\mathrm{sCM20}$ to vary. For this, we calculated a grid of $2.6 \leq \alpha_\mathrm{sCM20} \leq 5.4$ (where 5 is the THEMIS default) in steps of 0.01, for each value of the ISRF strength defined earlier (leading to a total grid of some 100,000 combination of parameters). The inclusion of fitting $\alpha_\mathrm{sCM20}$ brings our total number of free parameters to 6, and the best fit is shown, again, in Fig. \ref{fig:themis_sed_fit}. We also show the corner plot of this fit in Fig. \ref{fig:themis_corner}. We find a median $\alpha_\mathrm{sCM20}$ of around 4.3, somewhat lower than the THEMIS default of 5. In terms of the SED, this leads to a flatter slope at longer wavelengths. Physically, this corresponds to fewer very small carbon grains, as we might expect in a lower-metallicity environment such as M33. Much like the work of \cite{2017Chastenet} on the LMC and SMC, we find a much lower value for the silicate/carbon ratio of $\sim$0.3, compared to the MW value of $\sim$10. However, the ratio of small-to-large grains is very similar to the MW value of 0.4, with a value of 0.3.

Finally, we note that the fit does not perform so well in the 24-70\,\micron\, range. This can be improved by adding a second, warmer dust component (i.e. a higher ISRF strength). This produces a better fit in these wavelength ranges, but does not change the dust component masses, or $\alpha_\mathrm{sCM20}$ significantly. As we are only performing this fit to calculate the dust grain properties, and leave {\sc skirt} to model the ISRF, we only show the single-temperature component fit here.

\section{Rotating and Projecting the Data}\label{app:rotate_project_data}

\begin{figure*}
\centering
\includegraphics[width=2\columnwidth]{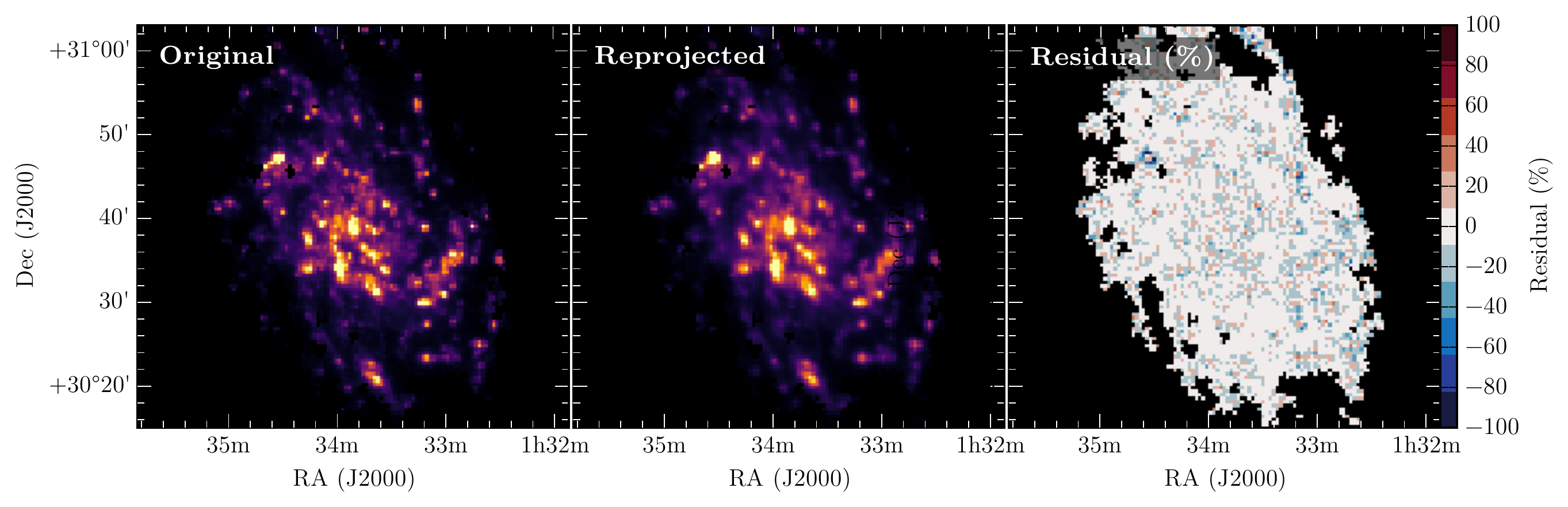}
\caption{The effect of deprojection and derotation on an input image. From left to right, we show the original GALEX FUV image, the image after derotation, deprojection, reprojection, and rotation, and the corresponding residuals of these two maps.}
\label{fig:appendix_derotate_deproject}
\end{figure*}

To add a 3D scale to the provided images, {\sc skirt} deprojects and de-rotates the input image, given an inclination and position angle. This means that the image becomes ``smeared'' as it is transformed into the plane of the galaxy, and then back into the observer frame. To test the effect that this has on our images (particularly for the purposes of residuals), we de-rotated and deprojected M33 (using a PA of 22.5\degr and inclination of 56\degr), before rotating and projecting it back into its original frame. The result of this can be seen in Fig. \ref{fig:appendix_derotate_deproject}. The effects of this routine are minor and will not affect our residual plots in any significant way, so we opt to use the original images as-is.


\bsp	
\label{lastpage}
\end{document}